\documentclass[11pt,letterpaper]{article}
\pdfoutput=1

\usepackage[utf8x]{inputenc}%
\usepackage{tcolorbox,fancybox}%
\usepackage{physics}
\usepackage{cite}
\usepackage{moresize}
\usepackage{ntheorem}
\usepackage{mathtools}
\usepackage{wasysym}

\usepackage[english]{babel}
\usepackage{amsmath,amssymb,amsfonts}
\usepackage{graphicx}
\usepackage{booktabs}
\usepackage{color,xcolor}
\usepackage[numbers,sort&compress]{natbib}

\usepackage{tikz}
\usepackage[compat=1.1.0]{tikz-feynman}
\usetikzlibrary{positioning}
\usetikzlibrary{decorations.text}
\usetikzlibrary{decorations.pathmorphing}

\usetikzlibrary{calc}
\usetikzlibrary{shapes.misc}
\tikzset{
       >=latex,
   photon/.style={decorate, decoration={snake}, draw=black, thick},
   fermionnoarrow/.style={draw=black, postaction={decorate}, thick},
   scalar/.style={draw=black, postaction={decorate}, decoration={markings,mark=at position .55 with {\arrow{>}}}, thick, dashed},
   scalarnoarrow/.style={draw=black, postaction={decorate},  thick, dashed},
   fermion/.style={draw=black, postaction={decorate},decoration={markings,mark=at position .55 with {\arrow{>}}}, thick},
   gluon/.style={decorate, draw=black, decoration={coil,amplitude=4pt, segment length=5pt}, thick},
   vertex/.style={draw,shape=circle,fill=black,minimum size=3pt,inner sep=0pt},
   fillvertex/.style={draw,shape=circle,fill=black,minimum size=5pt,inner sep=0pt},
   openvertex/.style={draw,shape=circle,minimum size=5pt,inner sep=0pt},
   blob/.style={draw=red,shape=circle,fill=red,minimum size=6pt,inner sep=0pt},
   redvertex/.style={draw=red,shape=circle,fill=red,minimum size=3pt,inner sep=0pt},
   cross/.style={cross out, draw=black,thick, minimum size=5pt, inner sep=0pt, outer sep=0pt}
}

\usepackage{physics}
\usepackage{orcidlink}
\usepackage{slashed}
\usepackage{mathrsfs}
\usepackage{enumerate}
\usepackage{mdframed}
\usepackage{url}
\usepackage[makeroom]{cancel}
\usepackage{geometry}

\usepackage{pifont}

\newtheorem*{thm-non}{Theorem}

\newtheorem*{define}[thm-non]{Definition}

\usepackage{hyperref} %Automatically links \label and \ref commands; Always load last
\hypersetup{
   colorlinks=true,       % false: boxed links; true: colored links
   linkcolor=red,          % color of internal links
   citecolor=blue,        % color of links to bibliography
   filecolor=magenta,      % color of file links
   urlcolor=purple           % color of external links
}
\usepackage[all]{hypcap} %Link navagates to top of figure instead of caption (below fig)

\def\beqn{\begin{eqnarray}}
\def\eeqn{\end{eqnarray}}
\def\beqs{\begin{subequations}}
\def\eeqs{\end{subequations}}
\def\beq{\begin{equation}}
\def\eeq{\end{equation}}
\def\ba{\begin{array}}
\def\ea{\end{array}}

\def\non{\nonumber\\}

\def\hf{\frac{1}{2}}
\def\[{\left[}
\def\]{\right]}
\def\({\left(}
\def\){\right)}
\newcommand\para{\paragraph{}}

\def\gG{\rm G}

\newcommand{\rep}[1]{\mathbf{#1}}
\newcommand{\repb}[1]{\mathbf{\overline{#1}}}

\def\Bc{\mathcal{B}}

\def\Dc{\mathcal{D}}
\def\Ec{\mathcal{E}}
\def\Fc{\mathcal{F}}
\def\Gc{\mathcal{G}}
\def\Hc{\mathcal{H}}

\def\Lc{\mathcal{L}}

\def\Nc{\mathcal{N}}
\def\Oc{\mathcal{O}}

\def\Qc{\mathcal{Q}}
\def\Rc{\mathcal{R}}

\def\Tc{\mathcal{T}}

\def\Xc{\mathcal{X}}

\def\DG{\mathfrak{D}}  \def\dG{\mathfrak{d}}
\def\EG{\mathfrak{E}}  \def\eG{\mathfrak{e}}
  
  \def\gG{\mathfrak{g}}
  \def\hG{\mathfrak{h}}

  \def\nG{\mathfrak{n}}
  \def\oG{\mathfrak{o}}

  \def\sG{\mathfrak{s}}
  
\def\UG{\mathfrak{U}}  \def\uG{\mathfrak{u}}

\title{
{\bf The unification in an $\widehat {\mathfrak{s}\mathfrak{u}}(8)_{ k_U = 1}$ affine Lie algebra} \\
\author{\large Ning Chen$^{\,\heartsuit}$\,\orcidlink{0000-0002-0032-9012}, Zhanpeng Hou$^{\,\spadesuit}$\,\orcidlink{0000-0002-6035-368X}, Zhaolong Teng$^{\,\clubsuit}$\,\orcidlink{0000-0002-7141-2331}}
\date{\small \it
$^\heartsuit \, ^\spadesuit\, ^\clubsuit$School of Physics, Nankai University, Tianjin, 300071, China \\
}
}

\begin{document}

\maketitle
\setlength{\parskip}{0.2ex}

\begin{abstract}
\bigskip
A flavor-unified theory based on the simple Lie algebra of ${\mathfrak{s}\mathfrak{u}}(8)$ was previously proposed to generate the observed three-generational Standard Model quark/lepton mass hierarchies and the Cabibbo-Kobayashi-Maskawa mixing pattern due to their non-universal symmetry properties.
A level-$1$ affine Lie algebra of $\widehat{ \mathfrak{s}\mathfrak{u} }(8)_{ k_U =1}$ with the ${\cal N}=1$ supersymmetric extension is found to unify three gauge couplings through the maximally symmetry breaking pattern.
\end{abstract}

\vspace{12cm}
{\emph{Emails:}\\  
$^{\,\heartsuit}$\url{chenning_symmetry@nankai.edu.cn},\\
$^{\,\spadesuit}$\url{houzhanpeng@mail.nankai.edu.cn},\\
$^{\,\clubsuit}$ \url{tengcl@mail.nankai.edu.cn}
}

\thispagestyle{empty}  
\newpage  

\setcounter{page}{1}  

\vspace{1.0cm}
\eject
\tableofcontents

%###################################################################
\section{Introduction}
\label{section:intro}
%###################################################################
%

%%%%%%%%%%%%%%%%%%%%%%%%%%%%
\subsection{Historical review and recent progresses}
%%%%%%%%%%%%%%%%%%%%%%%%%%%%

\para
Grand Unified Theories (GUTs) were proposed to not only unify all three Standard Model (SM) symmetries into one simple Lie algebra of either $\sG\uG(5)$~\cite{Georgi:1974sy} or $\sG\oG(10)$~\cite{Fritzsch:1974nn}, but to also unify all SM fermions into some anomaly-free irreps of the $\gG_U$.
The Georgi-Glashow $\sG\uG(5)$ GUT~\cite{Georgi:1974sy} and the Fritzsch-Minkowski $\sG\oG(10)$ GUT~\cite{Fritzsch:1974nn} contain the chiral fermions of $3\times \[ \repb{5_F} \oplus \rep{10_F} \]$ and $3\times \rep{16_F}$, respectively.
Neither framework can fully explain the three-generational SM fermion mass hierarchies and the Cabibbo-Kobayashi-Maskawa (CKM) mixing pattern~\cite{Cabibbo:1963yz,Kobayashi:1973fv} due to their trivially repetitive flavor structures.

\para
In a pioneering work by Georgi~\cite{Georgi:1979md}, an extension beyond the minimal simple Lie algebra of ${\sG\uG}(N >5)$ was proposed to embed three-generational SM fermions non-trivially.
In his original proposal, any anti-symmetric chiral fermion irrep can appear at most once with the anomaly-free condition.
Physically, this leads to a non-repetitive structure for both left-handed and right-handed components of chiral fermions based on an ${\sG\uG}(11)$ Lie algebra.
Given the current LHC measurements of one single SM Higgs boson so far, which have already confirmed the hierarchical Yukawa couplings of the third-generational SM quarks/leptons and displayed the evidence for the muon~\cite{CMS:2022dwd,ATLAS:2022vkf}, Georgi's original proposal is insightful and can be further refined with the concept of chiral \underline{ir}reducible \underline{a}nomaly-\underline{f}ree \underline{f}ermion \underline{s}ets (IRAFFSs)~\cite{Chen:2023qxi} as follows
\begin{define}\label{def:IRAFFS}

A chiral IRAFFS is a set of left-handed anti-symmetric fermions of $\sum_\Rc m_\Rc \, \Fc_L(\Rc)$, with $m_\Rc$ being the multiplicities of a particular fermion irrep of $\Rc$.
Obviously, the anomaly-free condition reads $\sum_\Rc m_\Rc \, {\rm Anom}(  \Fc_L(\Rc) ) =0$.
We also require the following conditions to be satisfied for a chiral IRAFFS:
\begin{itemize}

\item the \underline{g}reatest \underline{c}ommon \underline{d}ivisor (GCD) of the $\{ m_\Rc \}$ should satisfy that ${\rm GCD} \{  m_\Rc \} =1$;

\item the fermions in a chiral IRAFFS can no longer be removed, which would otherwise bring non-vanishing gauge anomalies;

\item there should not be any singlet, self-conjugate, adjoint fermions, or vectorial fermion pairs in a chiral IRAFFS.

\end{itemize}

\end{define}
A minimal ${\sG\uG}(8)$ theory~\cite{Barr:2008pn,Chen:2023qxi,Chen:2024cht} was thus proposed to include two following chiral IRAFFSs at the GUT scale
\beqn\label{eq:SU8_3gen_fermions}
\{ f_L \}_{  {\sG\uG}(8) }^{n_g=3}&=& \Big[ \repb{8_F}^\omega \oplus \rep{28_F} \Big] \bigoplus \Big[ \repb{8_F}^{ \dot \omega } \oplus \rep{56_F} \Big] \,,~ {\rm dim}_{ \mathbf{F}}= 156\,, \non
&& \Omega \equiv ( \omega \,, \dot \omega ) \,, ~ \omega = ( 3\,, {\rm IV}\,, {\rm V}\,, {\rm VI}) \,, ~  \dot \omega = (\dot 1\,, \dot 2\,, \dot {\rm VII}\,, \dot {\rm VIII}\,, \dot {\rm IX} ) \,,
\eeqn
with undotted/dotted indices for the $\repb{8_F}$'s in the rank-$2$ chiral IRAFFS and the rank-$3$ chiral IRAFFS~\footnote{The rank-$2$ and the rank-$3$ chiral IRAFFSs are named after the ${\sG\uG}(8)$ rank-$2$ and rank-$3$ anti-symmetric fermions of $A_2=\rep{28_F}$ and $A_3=\rep{56_F}$, respectively.}, respectively.
We also distinguish the heavy partner fermions and the SM fermions in terms of the Roman numbers and the Arabic numbers.
To count the SM generations by decomposing the ${\sG\uG}(8)$ fermion irreps into the ${\sG\uG}(5)$ irreps~\cite{Georgi:1979md}, one finds three identical SM $\repb{5_F}$'s and three distinctive $\rep{10_F}$'s~\cite{Barr:2008pn}.
Hence, it is sufficient to obtain three distinctive SM generations in the UV setup of the ${\sG\uG}(8)$ theory.

\begin{table}[htp]
\begin{center}
\begin{tabular}{c|cccc}
\hline\hline
Fermions &  $\repb{8_F}^\Omega$ &  $\rep{28_F}$  &  $\rep{56_F}$  &     \\[1mm]
\hline
$\widetilde{ {\rm U}}(1)_T$ &  $-3t$  &  $+2t$  & $+t$ &      \\[1mm]
$\widetilde{ {\rm U}}(1)_{\rm PQ}$ &  $p$  &  $q_2$  & $q_3$ &      \\[1mm]
\hline
Higgs  &  $\repb{8_H}_{\,, \omega }$  & $\repb{28_H}_{\,, \dot \omega }$   & $\rep{70_H}$ &  $\rep{63_H}$    \\[1mm]
\hline
$\widetilde{ {\rm U}}(1)_T$ &  $+t$  &  $+2t$   &  $-4t$  &  $0$    \\[1mm]
$\widetilde{ {\rm U}}(1)_{\rm PQ}$ &  $-(p+q_2)$  &  $-(p+q_3 )$  & $-2q_2$ &  $0$ \\[1mm]
\hline\hline
\end{tabular}
\end{center}
\caption{The non-anomalous $\widetilde{ {\rm U}}(1)_T$ charges and the anomalous global $\widetilde{ {\rm U}}(1)_{\rm PQ}$ charges for the ${\sG\uG}(8)$ fermions and Higgs fields.}
\label{tab:U1TU1PQ}
\end{table}%

\para
With the concept of the chiral IRAFFSs in Eq.~\eqref{eq:SU8_3gen_fermions}, one identifies the global symmetries of the chiral fermions in Eq.~\eqref{eq:SU8_3gen_fermions} to be
\beqn\label{eq:flavor_SU8}
\widetilde{ \Gc}_{\rm global} \[{\sG\uG}(8)\,, n_g=3 \]&=& \Big[ \widetilde{ {\rm SU}}(4)_\omega  \otimes \widetilde{ {\rm U}}(1)_{T_2}  \otimes \widetilde{ {\rm U}}(1)_{{\rm PQ}_2} \Big]  \bigotimes \Big[ \widetilde{ {\rm SU}}(5)_{\dot \omega } \otimes \widetilde{ {\rm U}}(1)_{T_3} \otimes \widetilde{ {\rm U}}(1)_{{\rm PQ}_3 }  \Big]  \,,
\eeqn
with the $\widetilde{ {\rm U}}(1)_{{\rm PQ}_i}$ being the anomalous global Peccei-Quinn symmetries~\cite{Peccei:1977hh}.
The most general gauge-invariant Yukawa couplings at least include the following renormalizable and non-renormalizable terms~\footnote{The term of $\rep{56_F} \rep{56_F} \rep{28_H}  + H.c.$ vanishes due to the anti-symmetric property~\cite{Barr:2008pn}. Instead, only a $d=5$ non-renormalizable term of $\frac{1}{ M_{\rm pl} } \rep{56_F}  \rep{56_F}  \repb{28_{H}}_{\,,\dot \omega }^\dag  \rep{63_{H}} $ is possible to generate masses for vectorlike fermions in the $\rep{56_F}$. Since it transforms as an $\widetilde{ {\rm SU}}(5)_{\dot \omega }$ vector and carries non-vanishing $\widetilde {\rm U}(1)_{\rm PQ}$ charge of $p+3q_3 \neq 0$ from Eq.~\eqref{eq:PQcharges_SU8}, it is only possible due to the gravitational effect.}
\beqn\label{eq:Yukawa_SU8}
-\Lc_Y&=& Y_\Bc  \repb{8_F}^\omega  \rep{28_F}  \repb{8_{H}}_{\,,\omega }  +  Y_\Tc \rep{28_F} \rep{28_F} \rep{70_H} \non
&+&  Y_\Dc \repb{8_F}^{\dot \omega  } \rep{56_F}  \repb{28_{H}}_{\,,\dot \omega }   + \frac{ c_4 }{ M_{\rm pl} } \rep{56_F}  \rep{56_F}  \repb{28_{H}}_{\,,\dot \omega }^\dag  \rep{63_{H}} + H.c.\,.
\eeqn
All renormalizable Yukawa couplings and the Wilson coefficient of the non-renormalizable term are expected to be $(Y_\Bc \,, Y_\Tc\,, Y_\Dc\,, c_4)\sim\Oc(1)$.
Accordingly, the non-anomalous global $\widetilde{ {\rm U}}(1)_T$ charges and the anomalous global $\widetilde{ {\rm U}}(1)_{\rm PQ}$ charges for fermions and Higgs fields are assigned in Tab.~\ref{tab:U1TU1PQ}, where the anomalous global $\widetilde{ {\rm U}}(1)_{\rm PQ}$ charges are assigned such that
\beqn\label{eq:PQcharges_SU8}
&&  p : q_2  \neq -3 : +2  \,, \quad p  : q_3 \neq -3  : +1  \,.
\eeqn

\para
The adjoint Higgs field of $\rep{63_H}$ in Eq.~\eqref{eq:Yukawa_SU8} will maximally break the symmetry as follows~\cite{Li:1973mq}
\beqn\label{eq:63H_VEV}
&& {\sG\uG}(8)\xhookrightarrow{ \langle  \rep{63_H}\rangle } \gG_{441}  \,, \quad \gG_{441}\equiv  {\sG\uG}(4)_{s} \oplus {\sG\uG} (4)_W \oplus  {\uG} (1)_{X_0 } \non
&& \langle  \rep{63_H}\rangle  = \frac{1}{ 4 } {\rm diag}(- \mathbb{I}_{4\times 4} \,, +\mathbb{I}_{4\times 4} ) v_U \,.
\eeqn
In the physical basis, the maximally symmetry breaking pattern defines the ${\uG}(1)_{X_0}$ charges for the ${\sG\uG}(8)$ fundamental representation as follows
\beqn\label{eq:X0charge}
\Xc_{0\,, {\tt phys}}( \rep{8} ) &\equiv& {\rm diag} ( \underbrace{ - \frac{1}{4}  \mathbb{I}_{4 \times 4}  }_{ \rep{4_s} }\,, \underbrace{ +\frac{1}{4} \mathbb{I}_{4 \times 4}   }_{ \rep{4_W} } )\,.
\eeqn
With the maximally symmetry breaking pattern in Eq.~\eqref{eq:63H_VEV}, the gauge coupling of $\alpha_{X_0}$ associated with the ${\uG}(1)_{X_0}$ charges is identical to the ${\uG}(1)_1$ gauge coupling of $\alpha_1$ obtained from the Cartan subalgebra~\cite{Chen:2024deo}.
Accordingly, several relevant ${\sG\uG}(8)$ irreps are decomposed as follows~\footnote{The decompositions of the conjugated irreps are neglected, since their conformal dimensions are identical to the original irreps.}
\beqs\label{eqs:SU8irrep_decomp}
\beqn
\rep{8}&\hookrightarrow& ( \rep{4} \,, \rep{1}\,,  -\frac{1}{4} ) \oplus  (\rep{1}\,, \rep{4}  \,,  +\frac{1}{4}) \,,\label{eq:8_decomp} \\[1mm]
%%%%%%%%%%%%%%%%%%%%%%%%%%%%%%%%%%%%%%%%%%%%%
\rep{28}&\hookrightarrow& ( \rep{6}\,, \rep{ 1} \,, - \frac{1}{2}) \oplus ( \rep{1}\,, \rep{ 6} \,, +\frac{1}{2}) \oplus  ( \rep{4}\,, \rep{4} \,,  0) \,,\label{eq:28_decomp} \\[1mm]
%%%%%%%%%%%%%%%%%%%%%%%%%%%%%%%%%%%%%%%%%%%%%
\rep{56}&\hookrightarrow& ( \rep{ 1}\,, \repb{4} \,, +\frac{3}{4}) \oplus  ( \repb{ 4}\,, \rep{1} \,, -\frac{3}{4}) \oplus ( \rep{ 4}\,, \rep{6} \,, +\frac{1}{4})  \oplus ( \rep{ 6}\,, \rep{4} \,, -\frac{1}{4}) \,,\label{eq:56_decomp}  \\[1mm]
%%%%%%%%%%%%%%%%%%%%%%%%%%%%%%%%%%%%%%%%%%%%%
\rep{70}&\hookrightarrow&  ( \rep{1} \,, \rep{1 } \,, -1 ) \oplus ( \rep{1} \,, \rep{1 } \,, +1 ) \oplus  ( \rep{4} \,, \repb{4} \,, +\frac{1}{2} )  \oplus  ( \repb{4} \,, \rep{4} \,, -\frac{1}{2} )   \oplus ( \rep{6 } \,, \rep{6 } \,, 0 )  \,,\label{eq:70_decomp}
\eeqn
\eeqs
in the physical basis.
Separately, we will also define the {\it algebraic basis} for the conformal embedding, where the ${\uG}(1)_{X_0}$ charges for the ${\sG\uG}(8)$ fundamental representation are integers
\beqn\label{eq:X0charge_alg}
\Xc_{0\,, {\tt alg}}( \rep{8} ) &\equiv& {\rm diag} ( \underbrace{ -   \mathbb{I}_{4 \times 4}  }_{ \rep{4_s} }\,, \underbrace{ + \mathbb{I}_{4 \times 4}   }_{ \rep{4_W} } )\,.
\eeqn
All ${\sG\uG}(8)$ irreps in Tab.~\ref{tab:U1TU1PQ} also carry integer-valued ${\uG}(1)_{X_0}$ charges in the algebraic basis as follows
\beqs\label{eqs:SU8irrep_decomp_alg}
\beqn
\rep{8}&\hookrightarrow& ( \rep{4} \,, \rep{1}\,,  - 1 ) \oplus  (\rep{1}\,, \rep{4}  \,,  +1 ) \,,\label{eq:8_decomp_alg} \\[1mm]
%%%%%%%%%%%%%%%%%%%%%%%%%%%%%%%%%%%%%%%%%%%%%
\rep{28}&\hookrightarrow& ( \rep{6}\,, \rep{ 1} \,, - 2 ) \oplus ( \rep{1}\,, \rep{ 6} \,, + 2 ) \oplus  ( \rep{4}\,, \rep{4} \,,  0) \,,\label{eq:28_decomp_alg} \\[1mm]
%%%%%%%%%%%%%%%%%%%%%%%%%%%%%%%%%%%%%%%%%%%%%
\rep{56}&\hookrightarrow& ( \rep{ 1}\,, \repb{4} \,, + 3  ) \oplus  ( \repb{ 4}\,, \rep{1} \,, - 3  ) \oplus ( \rep{ 4}\,, \rep{6} \,, + 1 )  \oplus ( \rep{ 6}\,, \rep{4} \,, - 1  ) \,,\label{eq:56_decomp_alg}  \\[1mm]
%%%%%%%%%%%%%%%%%%%%%%%%%%%%%%%%%%%%%%%%%%%%%
\rep{70}&\hookrightarrow&  ( \rep{1} \,, \rep{1 } \,, -4 ) \oplus ( \rep{1} \,, \rep{1 } \,, +4 ) \oplus  ( \rep{4} \,, \repb{4} \,, +  2 )  \oplus  ( \repb{4} \,, \rep{4} \,, - 2 )   \oplus ( \rep{6 } \,, \rep{6 } \,, 0 )  \,.\label{eq:70_decomp_alg}
\eeqn
\eeqs

\para
There can be three possible sequential symmetry breaking patterns~\footnote{The acronyms stand for the strong-weak-weak (SWW), weak-strong-weak (WSW), and weak-weak-strong (WWS) symmetry breaking patterns.}, namely,
\beqs\label{eqs:SU8_Patterns}
\beqn
{\rm SWW}~&:&~ {\sG\uG}(8) \hookrightarrow \gG_{441} \hookrightarrow  \gG_{341} \hookrightarrow \gG_{331}  \hookrightarrow \gG_{\rm SM} \,, \non
&& \gG_{341} \equiv {\sG\uG}(3)_{c} \oplus  {\sG\uG}(4)_W \oplus  {\uG}(1)_{X_1 }  \,, \quad  \gG_{331} \equiv {\sG\uG}(3)_{c} \oplus  {\sG\uG}(3)_W \oplus  {\uG}(1)_{X_2 } \,,  \non
&&  v_{441}\simeq 1.4 \times 10^{17 }\,{\rm GeV}  \,,~ v_{341} \simeq  4.8\times 10^{15} \,{\rm GeV} \,, ~ v_{331} \simeq 4.8\times 10^{13} \,{\rm GeV} \,, \label{eq:SU8_SWW}\\[1mm]
%%%%%%%%%%%%%%%%%%%%%%%%%%%%%%%%%%%%%%%%%%%%%
{\rm WSW}~&:&~ {\sG\uG}(8) \hookrightarrow \gG_{441} \hookrightarrow \gG_{431} \hookrightarrow \gG_{331} \hookrightarrow \gG_{\rm SM} \,, \non
&& \gG_{431} \equiv {\sG\uG}(4)_{s}  \oplus  {\sG\uG}(3)_W \oplus  {\uG}(1)_{X_1 } \,, \non
&& v_{441}\simeq 1.4 \times 10^{17 }\,{\rm GeV}  \,,~ v_{431} \simeq  4.8\times 10^{15} \,{\rm GeV} \,, ~ v_{331} \simeq 4.8\times 10^{13} \,{\rm GeV} \,, \\[1mm]
%%%%%%%%%%%%%%%%%%%%%%%%%%%%%%%%%%%%%%%%%%%%%
{\rm WWS}~&:&~ {\sG\uG}(8) \hookrightarrow \gG_{441} \hookrightarrow \gG_{431} \hookrightarrow \gG_{421} \hookrightarrow \gG_{\rm SM} \,, \non
&& \gG_{421} \equiv {\sG\uG}(4)_{s} \oplus {\sG\uG}(2)_W \oplus   {\uG}(1)_{X_2 }  \,, \quad  \gG_{\rm SM} \equiv  {\sG\uG}(3)_{c} \oplus {\sG\uG}(2)_W \oplus  {\uG}(1)_{Y } \,, \non
&&  v_{441}\simeq 1.4 \times 10^{17 }\,{\rm GeV}  \,,~ v_{431} \simeq  4.8\times 10^{15} \,{\rm GeV} \,, ~ v_{421} \simeq 1.1\times 10^{15} \,{\rm GeV} \,,
\eeqn
\eeqs
which have been analyzed in details in Refs.~\cite{Chen:2024cht,Chen:2024deo,Chen:2024yhb}.
Three intermediate symmetry breaking stages exist above the EW scale by counting the rank of the ${\rm SU}(8)$ group.
Through the analysis of the emergent non-anomalous global $\widetilde{ {\rm U}}(1)_T$ symmetry in Eq.~\eqref{eq:flavor_SU8}, we derived the non-anomalous global $B-L$ symmetry in the SM, and one unique SM Higgs doublet of $( \rep{1} \,, \repb{2} \,, +\frac{1}{2} )_{\mathbf{H}}^{\prime \prime\prime } \subset \rep{70_H}$ was conjectured~\cite{Chen:2023qxi} based on (i) its vanishing global $B-L$ charge, and (ii) its renormalizable Yukawa coupling to the top quark at the tree level.
Due to the distinctive symmetry properties of three-generational SM fermions, and with the unique SM Higgs doublet of $( \rep{1} \,, \repb{2} \,, +\frac{1}{2} )_{\mathbf{H}}^{\prime \prime\prime } \subset \rep{70_H}$ as the guidance, we further found in Ref.~\cite{Chen:2024cht} that all SM quark/lepton masses as well as the CKM mixing pattern can be explained with (i) both $d=5$ direct Yukawa coupling terms and indirect Yukawa coupling terms to the $d=5$ irreducible Higgs mixing operators induced by the inevitable gravitational effects that break the global flavor symmetries in Eq.~\eqref{eq:flavor_SU8}, (ii) three reasonable intermediate symmetry breaking scales suggested in Eqs.~\eqref{eqs:SU8_Patterns}, and (iii) precise identifications of the SM flavors in their UV irreps~\cite{Chen:2024cht,Chen:2024yhb}.
Specifically, the $d=5$ direct Yukawa coupling terms include
\beqn\label{eq:d5_direct}
c_4\, \Oc_{\Fc }^{(4\,,1)}&\equiv& c_4\, \rep{56_F}  \rep{56_F} \cdot \repb{28_H}_{\,, \dot \omega }  \cdot  \rep{70_H} \,,\non
c_5\, \Oc_{\Fc }^{(5\,,1)}&\equiv&  c_5\, \rep{28_F}  \rep{56_F} \cdot  \repb{8_H}_{\,,\omega }  \cdot  \rep{70_H}  \,.
\eeqn
%
%

%###############################################################################
\begin{table}[htp] {\small
\begin{center}
\begin{tabular}{c|c|c|c|c}
\hline \hline
  $\sG\uG(8)$   &  $\gG_{441}$  & $\gG_{341}$  & $\gG_{331}$  &  $\gG_{\rm SM}$  \\
\hline \hline
$\repb{ 8_F}^\Omega $   & $( \repb{4} \,, \rep{1}\,,  +\frac{1}{4} )_{ \mathbf{F} }^\Omega$  & $(\repb{3} \,, \rep{1} \,, +\frac{1}{3} )_{ \mathbf{F} }^\Omega $  & $(\repb{3} \,, \rep{1} \,, +\frac{1}{3} )_{ \mathbf{F} }^\Omega $  &  $( \repb{3} \,, \rep{ 1}  \,, +\frac{1}{3} )_{ \mathbf{F} }^{\Omega }~:~ { \Dc_R^\Omega }^c$  \\[1mm]
&  &  $( \rep{1} \,, \rep{1} \,, 0)_{ \mathbf{F} }^{\Omega }$  &  $( \rep{1} \,, \rep{1} \,, 0)_{ \mathbf{F} }^{\Omega }$ &  $( \rep{1} \,, \rep{1} \,, 0)_{ \mathbf{F} }^{\Omega } ~:~ \check \Nc_L^{\Omega }$  \\[1.5mm]
%%%%%%%%%%%%%%%%%%%%%%%%%%%%%%%%%%%%%%%%%%%%%
& $(\rep{1}\,, \repb{4}  \,,  -\frac{1}{4})_{ \mathbf{F} }^\Omega $  &  $(\rep{1}\,, \repb{4}  \,,  -\frac{1}{4})_{ \mathbf{F} }^\Omega$  &  $( \rep{1} \,, \repb{3} \,,  -\frac{1}{3})_{ \mathbf{F} }^{\Omega }$  &  $( \rep{1} \,, \repb{2} \,,  -\frac{1}{2})_{ \mathbf{F} }^{\Omega } ~:~\Lc_L^\Omega =( \Ec_L^\Omega \,, - \Nc_L^\Omega )^T$   \\[1mm]
&   &   &   &  $( \rep{1} \,, \rep{1} \,,  0)_{ \mathbf{F} }^{\Omega^\prime} ~:~ \check \Nc_L^{\Omega^\prime }$  \\[1mm]
 &   &  &   $( \rep{1} \,, \rep{1} \,, 0)_{ \mathbf{F} }^{\Omega^{\prime\prime} }$ &   $( \rep{1} \,, \rep{1} \,, 0)_{ \mathbf{F} }^{\Omega^{\prime\prime} } ~:~ \check \Nc_L^{\Omega^{\prime \prime} }$   \\[1mm]  
%%%%%%%%%%%%%%%%%%%%%%%%%%%%%%%%%%%%%%%%%%%%%
\hline\hline
\end{tabular}
\caption{
The $\sG\uG(8)$ fermion representation of $\repb{8_F}^\Omega$ under the $\gG_{441}\,,\gG_{341}\,, \gG_{331}\,, \gG_{\rm SM}$ subalgebras for the three-generational ${\sG\uG}(8)$ theory, with $\Omega\equiv (\omega \,, \dot \omega )$.
Here, we denote $\underline{ {\Dc_R^\Omega}^c={d_R^\Omega}^c}$ for the SM right-handed down-type quarks, and ${\Dc_R^\Omega}^c={\DG_R^\Omega}^c$ for the right-handed down-type heavy partner quarks.
Similarly, we denote $\underline{ \Lc_L^\Omega = ( \ell_L^\Omega \,, - \nu_L^\Omega)^T}$ for the left-handed SM lepton doublets, and $\Lc_L^\Omega =( \eG_L^\Omega \,, - \nG_L^\Omega )^T$ for the left-handed heavy lepton doublets.
All left-handed neutrinos of $\check \Nc_L$ are sterile neutrinos, which are $\Gc_{\rm SM}$-singlets and do not couple to the EW gauge bosons.
}
\label{tab:SU8_8barferm}
\end{center}
}
\end{table}%
%###############################################################################

%###############################################################################
\begin{table}[htp] {\small
\begin{center}
\begin{tabular}{c|c|c|c|c}
\hline \hline
  $\sG\uG(8)$   &  $\gG_{441}$  & $\gG_{341}$  & $\gG_{331}$  &  $\gG_{\rm SM}$  \\
\hline \hline
$\rep{28_F}$   & $( \rep{6}\,, \rep{ 1} \,, - \frac{1}{2})_{ \mathbf{F}}$ &  $ ( \rep{3}\,, \rep{ 1} \,, - \frac{1}{3})_{ \mathbf{F}}$   & $( \rep{3}\,, \rep{ 1} \,, - \frac{1}{3})_{ \mathbf{F}}$  & $( \rep{3}\,, \rep{ 1} \,, - \frac{1}{3})_{ \mathbf{F}} ~:~\DG_L$  \\[1mm]
                       &   & $( \repb{3}\,, \rep{ 1} \,, - \frac{2}{3})_{ \mathbf{F}}$  & $( \repb{3}\,, \rep{ 1} \,, - \frac{2}{3})_{ \mathbf{F}}$  & $\underline{( \repb{3}\,, \rep{ 1} \,, - \frac{2}{3})_{ \mathbf{F}}~:~ {t_R }^c }$   \\[1.5mm]
%%%%%%%%%%%%%%%%%%%%%%%%%%%%%%%%%%%%%%%%%%%%%
                       & $( \rep{1}\,, \rep{ 6} \,, +\frac{1}{2})_{ \mathbf{F}}$ & $( \rep{1}\,, \rep{ 6} \,, +\frac{1}{2})_{ \mathbf{F}}$   &  $( \rep{1}\,, \rep{ 3} \,, +\frac{1}{3})_{ \mathbf{F}}$ & $( \rep{1}\,, \rep{2} \,, +\frac{1}{2})_{ \mathbf{F}} ~:~( {\eG_R }^c \,, { \nG_R }^c)^T$  \\[1mm]
                      &   &   &   & $( \rep{1}\,, \rep{1} \,, 0 )_{ \mathbf{F}} ~:~ \check \nG_R^c $ \\[1mm]
                      &   &   & $( \rep{1}\,, \repb{ 3} \,, +\frac{2}{3})_{ \mathbf{F}}$  & $( \rep{1}\,, \repb{2} \,, +\frac{1}{2})_{ \mathbf{F}}^\prime ~:~( { \nG_R^{\prime} }^c\,, - {\eG_R^{\prime} }^c  )^T$   \\[1mm]
                      &   &   &   & $\underline{ ( \rep{1}\,, \rep{1} \,, +1 )_{ \mathbf{F}} ~:~ {\tau_R}^c}$ \\[1.5mm]
%%%%%%%%%%%%%%%%%%%%%%%%%%%%%%%%%%%%%%%%%%%%%
                       & $( \rep{4}\,, \rep{4} \,,  0)_{ \mathbf{F}}$ &  $( \rep{3}\,, \rep{4} \,,  -\frac{1}{12})_{ \mathbf{F}}$   & $( \rep{3}\,, \rep{3} \,,  0)_{ \mathbf{F}}$  & $\underline{ ( \rep{3}\,, \rep{2} \,,  +\frac{1}{6})_{ \mathbf{F}}~:~ (t_L\,, b_L)^T}$  \\[1mm]
                       &   &   &   & $( \rep{3}\,, \rep{1} \,,  -\frac{1}{3})_{ \mathbf{F}}^{\prime} ~:~\DG_L^\prime$  \\[1mm]
                       &   &   & $( \rep{3}\,, \rep{1} \,,  -\frac{1}{3})_{ \mathbf{F}}^{\prime\prime}$  & $( \rep{3}\,, \rep{1} \,,  -\frac{1}{3})_{ \mathbf{F}}^{\prime\prime} ~:~\DG_L^{\prime \prime}$ \\[1mm]
                       &   & $ ( \rep{1}\,, \rep{4} \,,  +\frac{1}{4} )_{ \mathbf{F}}$  & $( \rep{1}\,, \rep{3} \,,  +\frac{1}{3} )_{ \mathbf{F}}^{\prime\prime}$  & $( \rep{1}\,, \rep{2} \,,  +\frac{1}{2} )_{ \mathbf{F}}^{\prime\prime} ~:~( {\eG_R^{\prime\prime} }^c \,, { \nG_R^{\prime\prime}}^c )^T$  \\[1mm]
                       &   &   &   & $( \rep{1}\,, \rep{1}\,, 0)_{ \mathbf{F}}^{\prime} ~:~ \check \nG_R^{\prime\,c}$ \\[1mm]  
                       &   &   & $( \rep{1}\,, \rep{1}\,, 0)_{ \mathbf{F}}^{\prime\prime}$ & $( \rep{1}\,, \rep{1}\,, 0)_{ \mathbf{F}}^{\prime\prime} ~:~\check \nG_R^{\prime \prime \,c}$ \\[1mm]  
%%%%%%%%%%%%%%%%%%%%%%%%%%%%%%%%%%%%%%%%%%%%%
\hline\hline
\end{tabular}
\caption{
The $\sG\uG(8)$ fermion representation of $\rep{28_F}$ under the $\gG_{441}\,,\gG_{341}\,, \gG_{331}\,, \gG_{\rm SM}$ subalgebras for the three-generational ${\sG\uG}(8)$ theory.
All SM fermions are marked with underlines.}
\label{tab:SU8_28ferm}
\end{center}
}
\end{table}%
%###############################################################################

%###############################################################################
\begin{table}[htp] {\small
\begin{center}
\begin{tabular}{c|c|c|c|c}
\hline \hline
  $\sG\uG(8)$   &  $\gG_{441}$  & $\gG_{341}$  & $\gG_{331}$  &  $\gG_{\rm SM}$  \\
\hline \hline
    $\rep{56_F}$   & $( \rep{ 1}\,, \repb{4} \,, +\frac{3}{4})_{ \mathbf{F}}$  &  $( \rep{ 1}\,, \repb{4} \,, +\frac{3}{4})_{ \mathbf{F}}$ & $( \rep{ 1}\,, \repb{3} \,, +\frac{2}{3})_{ \mathbf{F}}^\prime$   &  $( \rep{ 1}\,, \repb{2} \,, +\frac{1}{2})_{ \mathbf{F}}^{\prime\prime\prime} ~:~( {\nG_R^{\prime\prime\prime }}^c \,, -{\eG_R^{\prime\prime\prime } }^c )^T$  \\[1mm]
                           &   &   &   & $\underline{( \rep{ 1}\,, \rep{1} \,, +1)_{ \mathbf{F}}^{\prime} ~:~ {\mu_R}^c}$ \\[1mm]
                           &   &   & $( \rep{ 1}\,, \rep{1} \,, +1)_{ \mathbf{F}}^{\prime\prime}$  & $( \rep{ 1}\,, \rep{1} \,, +1)_{ \mathbf{F}}^{\prime \prime} ~:~{\EG_R}^c$   \\[1.5mm]
%%%%%%%%%%%%%%%%%%%%%%%%%%%%%%%%%%%%%%%%%%%%%
                      & $( \repb{ 4}\,, \rep{1} \,, -\frac{3}{4})_{ \mathbf{F}}$  &  $( \repb{3}\,, \rep{1} \,, -\frac{2}{3})_{ \mathbf{F}}^{\prime}$ & $( \repb{3}\,, \rep{1} \,, -\frac{2}{3})_{ \mathbf{F}}^\prime$  & $\underline{ ( \repb{3}\,, \rep{1} \,, -\frac{2}{3})_{ \mathbf{F}}^{\prime} ~:~{u_R}^c }$ \\[1mm]
                      &   &  $( \rep{1}\,, \rep{1} \,, -1)_{ \mathbf{F}}$ & $( \rep{1}\,, \rep{1} \,, -1)_{ \mathbf{F}}$  &  $( \rep{1}\,, \rep{1} \,, -1)_{ \mathbf{F}} ~:~\EG_L$  \\[1.5mm]
%%%%%%%%%%%%%%%%%%%%%%%%%%%%%%%%%%%%%%%%%%%%%
                      & $( \rep{ 4}\,, \rep{6} \,, +\frac{1}{4})_{ \mathbf{F}}$  &  $( \rep{3}\,, \rep{6} \,, +\frac{1}{6})_{ \mathbf{F}}$ & $( \rep{3}\,, \rep{3} \,, 0 )_{ \mathbf{F}}^\prime$ & $\underline{ ( \rep{3}\,, \rep{2} \,, +\frac{1}{6} )_{ \mathbf{F}}^{\prime} ~:~ ( c_L \,, s_L)^T} $  \\[1mm]
                      &   &   &   & $( \rep{3}\,, \rep{1} \,, -\frac{1}{3})_{ \mathbf{F}}^{\prime\prime \prime } ~:~\DG_L^{\prime \prime \prime}$ \\[1mm]
                      &   &   & $( \rep{3}\,, \repb{3} \,, +\frac{1}{3})_{ \mathbf{F}}$ & $( \rep{3}\,, \repb{2} \,, +\frac{1}{6})_{ \mathbf{F}}^{\prime\prime} ~:~ (\dG_L \,, - \uG_L )^T$   \\[1mm]
                      &   &   &   & $( \rep{3}\,, \rep{1} \,, +\frac{2}{3})_{ \mathbf{F}} ~:~\UG_L$  \\[1mm]
                      &   & $( \rep{1}\,, \rep{6} \,, +\frac{1}{2})_{ \mathbf{F}}^\prime$ & $( \rep{1}\,, \rep{3} \,, +\frac{1}{3})_{ \mathbf{F}}^\prime $ & $( \rep{1}\,, \rep{2} \,, +\frac{1}{2})_{ \mathbf{F}}^{\prime\prime \prime \prime} ~:~ ( {\eG_R^{\prime\prime\prime\prime }}^c \,, {\nG_R^{\prime\prime\prime\prime } }^c )^T$ \\[1mm]
                      &   &   &   & $( \rep{1}\,, \rep{1} \,, 0 )_{ \mathbf{F}}^{\prime\prime \prime} ~:~ {\check \nG_R}^{\prime \prime\prime \,c}$ \\[1mm]
                      &   &   & $( \rep{1}\,, \repb{3} \,, +\frac{2}{3})_{ \mathbf{F}}^{\prime\prime}$  & $( \rep{1}\,, \repb{2} \,, +\frac{1}{2})_{ \mathbf{F}}^{\prime\prime \prime \prime \prime} ~:~( {\nG_R^{\prime\prime\prime\prime\prime }}^c \,, -{\eG_R^{\prime\prime\prime\prime\prime } }^c )^T$  \\[1mm]
                      &   &   &   & $\underline{ ( \rep{1}\,, \rep{1} \,, +1 )_{ \mathbf{F}}^{\prime\prime \prime } ~:~ {e_R}^c }$ \\[1.5mm]
%%%%%%%%%%%%%%%%%%%%%%%%%%%%%%%%%%%%%%%%%%%%%
                      & $( \rep{ 6}\,, \rep{4} \,, -\frac{1}{4})_{ \mathbf{F}}$  & $( \rep{3}\,, \rep{4} \,, -\frac{1}{12})_{ \mathbf{F}}^\prime$ & $( \rep{3}\,, \rep{3} \,, 0)_{ \mathbf{F}}^{\prime\prime}$ & $\underline{ ( \rep{3}\,, \rep{2} \,, +\frac{1}{6})_{ \mathbf{F}}^{\prime\prime \prime } ~:~  (u_L \,, d_L)^T} $ \\[1mm]
                      &   &   &   &  $( \rep{3}\,, \rep{1} \,, -\frac{1}{3})_{ \mathbf{F}}^{\prime \prime \prime \prime} ~:~\DG_L^{\prime \prime \prime\prime}$ \\[1mm]
                      &   &   &  $( \rep{3}\,, \rep{1} \,, -\frac{1}{3})_{ \mathbf{F}}^{\prime \prime \prime\prime \prime}$ & $( \rep{3}\,, \rep{1} \,, -\frac{1}{3})_{ \mathbf{F}}^{\prime \prime \prime \prime \prime} ~:~ \DG_L^{\prime \prime \prime\prime \prime}$ \\[1mm]
                      &   & $( \repb{3}\,, \rep{4} \,, -\frac{5}{12})_{ \mathbf{F}}$ & $( \repb{3}\,, \rep{3} \,, -\frac{1}{3})_{ \mathbf{F}}$ & $( \repb{3}\,, \rep{2} \,, -\frac{1}{6})_{ \mathbf{F}} ~:~ ( {\dG_R}^c \,,{\uG_R}^c )^T$  \\[1mm]
                      &   &   &   & $( \repb{3}\,, \rep{1} \,, -\frac{2}{3})_{ \mathbf{F}}^{\prime \prime} ~:~{\UG_R}^c$  \\[1mm]
                      &   &   & $( \repb{3}\,, \rep{1} \,, -\frac{2}{3})_{ \mathbf{F}}^{\prime \prime \prime}$ & $\underline{ ( \repb{3}\,, \rep{1} \,, -\frac{2}{3})_{ \mathbf{F}}^{\prime \prime \prime} ~:~{c_R}^c }$  \\[1mm]
%%%%%%%%%%%%%%%%%%%%%%%%%%%%%%%%%%%%%%%%%%%%%
\hline\hline
\end{tabular}
\caption{
The $\sG\uG(8)$ fermion representation of $\rep{56_F}$ under the $\gG_{441}\,,\gG_{341}\,, \gG_{331}\,, \gG_{\rm SM}$ subalgebras for the three-generational ${\sG\uG}(8)$ theory.
All SM fermions are marked with underlines.
}
\label{tab:SU8_56ferm}
\end{center}
}
\end{table}%
%###############################################################################

\para
For the later convenience, we shall focus on the SWW symmetry breaking pattern in Eq.~\eqref{eq:SU8_SWW} throughout this paper.
Accordingly, we tabulate the fermion representations at various stages of the ${\rm SU}(8)$ model in Tabs.~\ref{tab:SU8_8barferm}, \ref{tab:SU8_28ferm}, and \ref{tab:SU8_56ferm}.

%%%%%%%%%%%%%%%%%%%%%%%%%%%%
\subsection{The main results}
%%%%%%%%%%%%%%%%%%%%%%%%%%%%

%%%%%%%%%%%%%%%%%%%%%%%%%%%%%%%%%%%%%%%%%%%%%%%%%%%%%%%
\begin{figure}[htb]
\centering
\includegraphics[height=6.5cm]{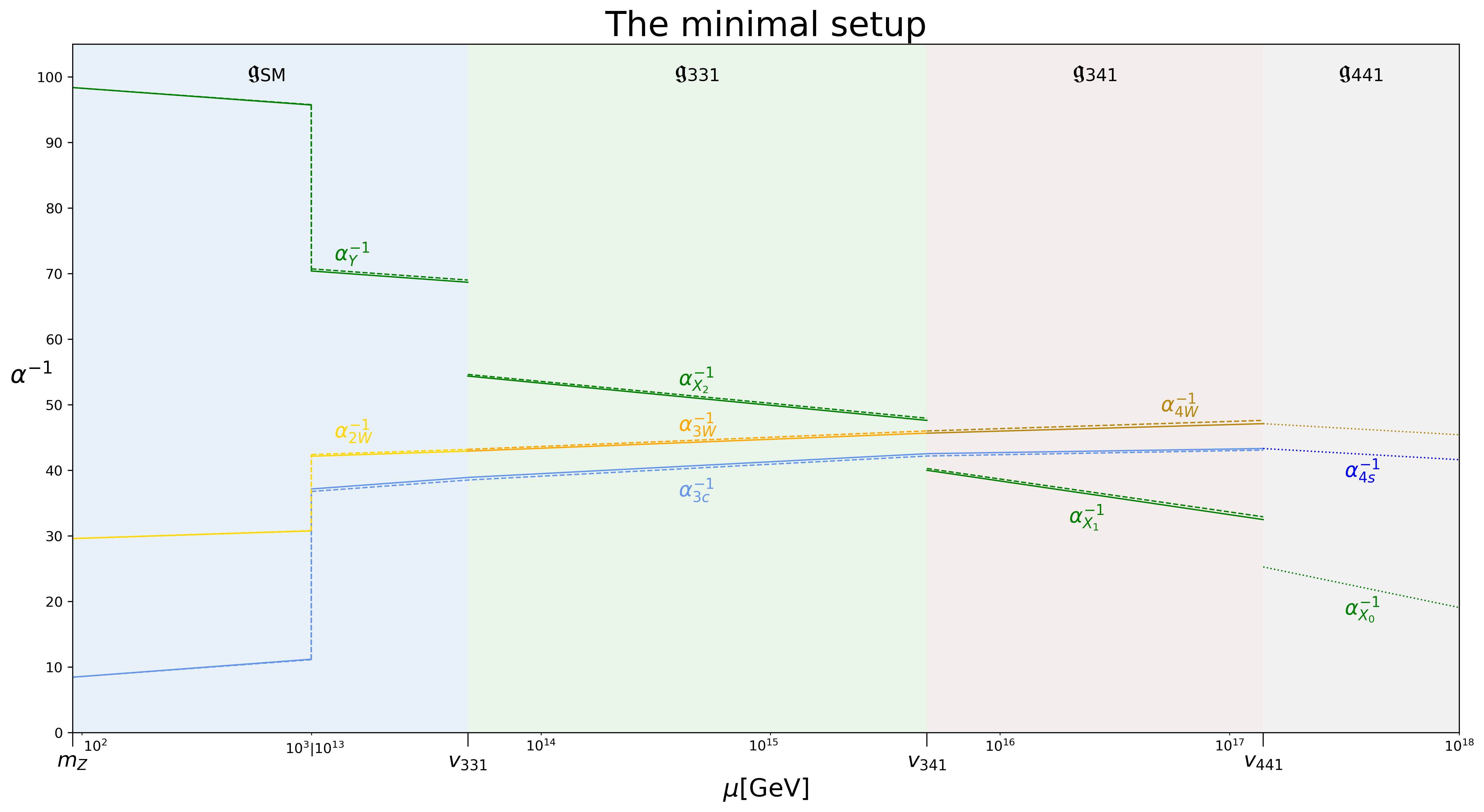}
\caption{The RGEs of the minimal non-SUSY ${\sG\uG}(8)$ setup through the SWW symmetry breaking pattern in Eq.~\eqref{eq:SU8_SWW}.
The dashed lines and the solid lines represent the one-loop and two-loop RGEs, respectively.
The interval of $10^{3}\,{\rm GeV} \lesssim \mu \lesssim 10^{13}\,{\rm GeV}$ is zoomed out in order to highlight the behaviors in three intermediate symmetry breaking scales.
}
\label{fig:RGE_mini}
\end{figure}
%%%%%%%%%%%%%%%%%%%%%%%%%%%%%%%%%%%%%%%%%%%%%%%%%%%%%%%

\para
The GUT can only be valid when the gauge coupling unification is achieved in terms of their renormalization group equations (RGEs)~\footnote{Some early studies of the RGEs based on the ${\sG\uG}(5)$ and ${\sG\oG}(10)$ groups include Refs.~\cite{Georgi:1974yf,Hall:1980kf,Dimopoulos:1981zb}.}.
Based on the extended gauge sectors and the suggested intermediate symmetry breaking scales in Eqs.~\eqref{eqs:SU8_Patterns}, we perform the analyses of the gauge coupling RGEs in Refs.~\cite{Chen:2024deo,Chen:2024yhb}.
It turns out that the gauge couplings cannot achieve the unification in the minimal ${ \sG\uG}(8)$ setup, with effects of (i) different hypotheses of the Higgs fields' masses in the spectrum, (ii) the reasonable threshold effects~\cite{Weinberg:1980wa,Hall:1980kf}, and (iii) the possible $d=5$ gravity-induced operators of
\beqn\label{eq:HSW_Op}
\Oc_{\rm HSW}&\equiv& - \hf \frac{ c_{\rm HSW} }{ M_{\rm pl}} \Tr[ \rep{63_H} U^{ \mu \nu } U_{\mu \nu } ] \,,
\eeqn
suggested by Hill-Shafi-Wetterich~\cite{Hill:1983xh,Shafi:1983gz} taken into account.
Here, the $U_{\mu\nu}$ represents the ${\sG\uG}(8)$ field strength tensor.
After the GUT-scale symmetry breaking with the VEV in Eq.~\eqref{eq:63H_VEV}, this HSW operator shifts two ${\sG\uG}(4)_{ s /W}$ field strengths of $G_{ \mu \nu }=G_{ \mu \nu }^{\bar A} T_{ {\rm SU}(4)_s }^{\bar A}$ and $W_{ \mu \nu }= W_{ \mu \nu }^{\bar I} T_{  {\rm SU}(4)_W }^{\bar I}$ as follows
\beqn
&& - \frac{1 }{2 } ( 1 - \frac{ 1  }{4  } c_{\rm HSW} \zeta_0 ) \Tr ( G_{ \mu \nu } G^{ \mu \nu } ) - \frac{1 }{2 } ( 1 + \frac{ 1  }{4  } c_{\rm HSW} \zeta_0 ) \Tr (  W_{ \mu \nu }  W^{ \mu \nu } )\,,
\eeqn
with $\zeta_0\equiv v_U/ M_{\rm pl}$.
The non-Abelian gauge coupling unification is modified into
\beqn\label{eq:unification_HSW}
&& ( 1 - \frac{ 1  }{4  } c_{\rm HSW} \zeta_0 ) \alpha_{4 s} ( v_U ) =  ( 1 + \frac{ 1  }{4  } c_{\rm HSW} \zeta_0 ) \alpha_{4W} ( v_U )   = \alpha_{U } ( v_U ) \,,
\eeqn
while the Abelian gauge coupling is untouched with the maximally symmetry breaking pattern in Eq.~\eqref{eq:63H_VEV}.
A typical example of the RGEs in the minimal ${\rm SU}(8)$ setup is given in Fig.~\ref{fig:RGE_mini}, with the survival hypothesis~\cite{Georgi:1979md,Glashow:1979nm,Barbieri:1979ag,Barbieri:1980vc,Barbieri:1980tz,delAguila:1980qag} to the Higgs spectrum~\footnote{An alternative assumption to the minimal ${\sG\uG}(8)$ Higgs spectrum was also considered in Ref.~\cite{Chen:2024deo} and lead to a very similar behavior of the RGEs.}.
In other words, the difficulty of achieving the gauge coupling unification mainly comes from: (i) the defined {\it Cartan discontinuities} of ${\uG}(1)$ gauge couplings in Eq.~\eqref{eq:Cartan_discont}, and (ii) the suggested intermediate symmetry breaking scales, which were previously determined according to the observed SM quark/lepton masses and the CKM mixing patterns~\cite{Chen:2024cht}.

\para
Furthermore, the previous analyses of different symmetry breaking patterns~\cite{Chen:2024yhb} seem to suggest a following gauge coupling relation of
\beqn
&& \alpha_{4 s}^{-1} (v_U) \approx \alpha_{4 W}^{-1} (v_U) \approx 2 \alpha_{X_0}^{-1} (v_U) = 40\,,\quad v_U\approx 5.0\times 10^{17}\,{\rm GeV}\,.
\eeqn
With the reasonable threshold effects taken into account, we have found that such a large discrepancy between the non-Abelian and the Abelian gauge couplings cannot be eliminated.
It was first shown by Ginsparg~\cite{Ginsparg:1987ee} that the gauge coupling unification in the string theory can be achieved as
%The above results motivate us to interpret the unification in the framework of the affine Lie algebra, which reads
%
%
\beqn\label{eq:affine_unification}
&& k_s \alpha_{4 s} (v_U) = k_W \alpha_{4W} (v_U) = k_1 \alpha_{ 1 }  (v_U)=  \alpha_{ U}  (v_U) \,,
\eeqn
with $(k_s\,, k_W\,, k_1)$ being the affine levels of the corresponding gauge couplings in the physical basis~\footnote{Some early studies of the higher affine levels in the string-inspired unified theories include Refs.~\cite{Font:1990uw,Dienes:1996du}.}.
In this paper, we prove that the reasonable conformal embedding is based on the affine Lie algebra of
\beqn\label{eq:SU8_embedding}
&& \widehat{ \sG \uG}(4)_{s\,, k_s = 1} \oplus \widehat{ \sG \uG}(4)_{W\,, k_W = 1} \oplus \hat{  \uG}(1)_{ 1 \,, k_{1\,,{\tt phys}} =  \frac{1}{4 } }  \subset \widehat{ \sG \uG}(8)_{ k_U = 1} \,,
\eeqn
in the {\it physical basis}, and the gauge coupling unification can be achieved with the $\Nc=1$ supersymmetric (SUSY) extension between the scale of $v_{441} \leq \mu \leq v_U$.

\para
The rest of the paper is organized as follows.
The conformal embedding of the affine Lie algebra in Eq.~\eqref{eq:SU8_embedding} will be derived in Sec.~\ref{section:constraints}.
In Sec.~\ref{section:SUSY_SU8}, we propose an $\Nc=1$ SUSY extension to the ${\sG\uG}(8)$ theory, and study the corresponding RGEs between the scale of $v_{441} \leq \mu \leq v_U$.
Such type of SUSY extension achieves the gauge coupling unification according to the corresponding affine Lie algebras, with the reasonable assumptions of the mass spectra.
We summarize the results and discuss the underlying issues in Sec.~\ref{section:discussions}.
In App.~\ref{section:affine}, we provide the necessary results of affine Lie algebra, with the emphasis on the root and weight system defined on the Cartan-Weyl basis.
In App.~\ref{section:discont}, we provide the general results of the ${\uG}(1)$ Cartan discontinuities at different symmetry breaking stages.

%###################################################################
\section{The conformal embedding of the $\widehat{{ \sG\uG}}(8)_{k_U}$ affine Lie algebra}
\label{section:constraints}
%###################################################################

\para
Throughout this section, we shall discuss the affine embeddings of the form $\oplus_{i} \hat \hG_{ i\,, k_i } \subset \hat \gG_k$, where $\hat \hG_{ i\,, k_i }$ and $\hat \gG_k$ represent some subalgebras and parent algebra, respectively.
Mostly, we will use the group invariants defined with the normalization of $T( \Box)_{\tt alg}=1$ in what we called the algebraic basis, and the corresponding quantities are denoted with a subscript of {\tt alg}.
The central issue is to look for the conformal embedding of the $\widehat{{ \sG\uG}}(8)_{k_U}$ affine Lie algebra according to the maximally symmetry breaking pattern in Eq~\eqref{eq:63H_VEV}.
Some well-known constraints to the $\widehat{ \sG\uG}(8)_{k_U}$ affine Lie algebra are
\begin{enumerate}

\item $k_U\in \mathbb{Z}^+$,

\item the central charge should satisfy $c(\widehat{ \sG\uG}(8)_{k_U}) \leq 22$, with the central charge defined by
\beqn
c(\hat{ \gG }_{k})&\equiv& \frac{k\, {\rm dim}(\gG ) }{ k + g^\lor }\,,
\eeqn
for an affine Lie algebra of $\hat{ \gG }_{k}$.
With the dual Coxeter number of $g^\lor=N$ for $\gG=\sG\uG(N)$ according to Eq.~\eqref{eq:dualCoxeter}, there is an upper bound to the $\widehat{ \sG\uG}(8)_{k_U}$ level as
\beqn
\widehat{ \sG\uG}(8)_{k_U}~&:&~ k_U \leq \frac{176}{41}\,,\quad {\rm or}~~ k_U \leq 4\,.
\eeqn

\end{enumerate}

\para
The conformal embedding is a subclass of the affine embeddings of $\oplus_{i} \hat \hG_{ i\,, k_i } \subset \hat \gG_k$ that preserves the conformal invariance, which can be satisfied when the corresponding central charges are equal
%For $\hat \pG_{ k^\prime } = \otimes \hat \hG_{ i\,, k_i }$, the above relation becomes
%
%
\beqn\label{eq:cc_relation}
&& \sum_i c( \hat \hG_{ i\, k_i } ) = c ( \hat \gG_k )  \Rightarrow  \sum_i  \frac{  k_i \, {\rm dim} \hG_i }{  k_i + h_i^\lor } = \frac{ k \, {\rm dim} \gG }{ k + g^\lor }  \,,
\eeqn
with $(h_i^\lor \,, g^\lor )$ being the dual Coxeter numbers of $\hG_i$ and $\gG$.
For the Abelian $\hat\uG(1)_{k_1}$, the central charge is reduced to
\beqn
&& c ( \hat\uG(1)_{k_1 } )= \frac{ k_1 {\rm dim} \uG(1) }{ k_1 } =1\,,
\eeqn
since the $\uG(1)$ has no dual Coxeter number.
It is known that the conformal embeddings are only possible for $k=1$.
For example, the $\widehat{ \sG \uG}(2)_1 \subset \widehat{ \sG \uG}(3)_1$ cannot be a conformal embedding, since the central charges are not equal
\beqn
&&c( \widehat{ \sG \uG}(2)_1) =  \frac{1 \times 3}{ 1+ 2}=1 \,, \quad c(\widehat{ \sG \uG}(3)_1 ) =  \frac{1\times 8 }{ 1 + 3} =2 \Rightarrow c( \widehat{ \sG \uG}(2)_1 ) \neq c ( \widehat{ \sG \uG}(3)_1) \,.
\eeqn
Instead, the $\widehat{ \sG \uG}(2)_4 \subset \widehat{ \sG \uG}(3)_1$ can be a conformal embedding, since
\beqn
&& c( \widehat{ \sG \uG}(2)_4) =  \frac{4 \times 3}{ 4+ 2}=2 \,, \quad c(\widehat{ \sG \uG}(3)_1 ) =  \frac{1\times 8 }{ 1 + 3} =2  \Rightarrow  c( \widehat{ \sG \uG}(2)_4 ) = c ( \widehat{ \sG \uG}(3)_1)  \,.
\eeqn

\para
With these relations, let us turn to our flavor-unified ${\rm SU}(8)$ theory with the maximally breaking pattern in Eq.~\eqref{eq:63H_VEV}.
If one has to achieve the conformal embedding of $\widehat{ \sG \uG}(4)_{s\,,k_s } \oplus \widehat{ \sG \uG}(4)_{W\,, k_W} \oplus \hat{  \uG}(1)_{ 1 \,, k_1}  \subset \widehat{ \sG \uG}( 8)_{ k_U = 1}$, the conformal invariance in Eq.~\eqref{eq:cc_relation} fixes the affine levels of both $\widehat{ \sG \uG}(4)_{s} \oplus \widehat{ \sG \uG}(4)_{W}$ subalgebras to be
\beqn
&&  \frac{15\, k_s }{k_s + 4} + \frac{15\, k_W }{k_W + 4} +1 =7 \Rightarrow ( k_s \,, k_W )=( 1 \,, 1) \,.
\eeqn
According to the relation in Eq.~\eqref{eq:affine_unification}, the corresponding affine levels of two non-Abelian Lie algebras are consistent with the relation of $\alpha_{4s}^{-1} (v_U )\approx \alpha_{4W}^{-1}(v_U)$ from the RGEs in Fig.~\ref{fig:RGE_mini}.
If one considers the generalized conformal embedding of the form $\widehat{ \sG \uG}( n_s )_{s\,,k_s } \oplus \widehat{ \sG \uG}( n_W )_{W\,, k_W} \oplus \hat{  \uG}(1)_{ 1 \,, k_1 }  \subset \widehat{ \sG \uG}( N)_{1}$ with $N=n_s + n_W$, one finds that the following equality
\beqn
&&  \frac{(n_s^2 -1)\, k_s }{k_s + n_s } + \frac{(n_W^2 -1)\, k_W }{k_W + n_W} +1 = N-1
\eeqn
between the central charges always holds when $( k_s \,, k_W )=( 1 \,, 1)$.
Meanwhile, the level of $\hat{  \uG}(1)_{k_1 } $ is undetermined by the conformal invariance.
Alternatively, if one proposes a different conformal embedding of $\widehat{ \sG \uG}(4)_{2 } \otimes \widehat{ \sG \uG}(4)_{2 } \otimes \hat{  \uG}(1)_{ k_1 }  \subset \widehat{ \sG \uG}( N)_{1}$, the relation in Eq.~\eqref{eq:cc_relation} leads to
\beqn
&& 2\times \frac{15 \times 2 }{2 +4} + 1 = N-1 \Rightarrow N = 12\,.
\eeqn
This means an embedding with higher affine levels can only be achieved through $\widehat{ \sG \uG}(4)_{2 } \oplus \widehat{ \sG \uG}(4)_{2 } \oplus \hat{  \uG}(1)_{ k_1}  \subset \widehat{ \sG \uG}( 12)_{1}$, which is inconsistent with our minimal setup described in Sec.~\ref{section:intro}.

\para
The unitarity constraint~\cite{Goddard:1986bp} further requires that the conformal dimension of the massless and highest-weight states should satisfy 
\beqn\label{eq:conformal_unitarity}
&& h(\Rc) \leq 1 \,,
\eeqn
where the conformal dimensions of the highest-weight states are given by
\beqn\label{eq:conformal_dim_nonAbelian}
h(\Rc)&\equiv& \frac{ C_2(\Rc )_{\tt alg} /( \boldsymbol{\alpha }_h \,, \boldsymbol{\alpha }_h ) }{ k_U + g_U^\lor } \,, % \frac{ C_2(\Rc ) / | \vec \alpha|^2 }{ k_U + g_U } \leq 1\,,
\eeqn
with $g_U^\lor=8$ for $\gG=\sG\uG(8)$, and $( \boldsymbol{\alpha }_h \,, \boldsymbol{\alpha }_h )=2$ according to the normalization in the algebraic basis~\footnote{Here, the $\boldsymbol{\alpha }_h$ represents the longest root of the Lie algebra $\gG$. For the $\sG\uG(N)$ Lie algebra, the length of all equally-long roots is normalized as $( \boldsymbol{\alpha }_h \,,\boldsymbol{\alpha }_h )=2$ in the algebraic basis.}.
More generically, the conformal dimensions of the rank-$k$ anti-symmetric irreps in the $\widehat{ \sG\uG}(N)_{ k_U = 1}$ theories are
\beqn
&& h (   A_k  )= \frac{ k (N - k) }{ 2 N }\,, %\quad h (  S_2  )= \frac{ ( N-1)( N+2) }{  N (N+1) } \quad \,,
%h(  {\rm adj}  ) = \frac{N }{1 +N }\,,
\eeqn
according to Eq.~\eqref{eq:Casimir-Ak}.
%Obviously, both the rank-$2$ symmetric~\footnote{Here, the rank-$2$ symmetric irrep of the $\rep{36}$ was not present in the minimal ${\rm SU}(8)$ theory, while it can be possible in a Type-II SUSY extension in Eq.~\eqref{eq:SUSY_SU8_chiralsII}. } and the adjoint irreps can satisfy the unitarity constraint in Eq.~\eqref{eq:conformal_unitarity}.
Specifically with $C_2(\rep{8})_{\tt alg} = C_2(\repb{8})_{\tt alg} =63/8$, $C_2(\rep{28})_{\tt alg} = C_2(\repb{28})_{\tt alg} =27/2$, $C_2(\rep{56})_{\tt alg}=135/8$, and $C_2(\rep{70})_{\tt alg}=18$ according to Eqs.~\eqref{eq:Casimir-Ak}, we find the following conformal dimensions
\beqn
&& h( \rep{8}) = h( \repb{8}) =  \frac{7}{16}\,, \quad h( \rep{28}) = h( \repb{28}) = \frac{ 3 }{ 4} \,, \quad  h( \rep{ 56}) = \frac{15}{16} \,, \quad h( \rep{70}) = 1 \,,
\eeqn
in the $\widehat{ \sG \uG}(8)_{k_U =1}$ theory. 
This means all massless states with the anti-symmetric irreps satisfy the unitarity constraint.

\para
Next, we also need to match the conformal dimensions of the fields through the affine branching rules of
\beqn
&& \hat{\boldsymbol{\lambda } }  \mapsto \bigoplus_{ \hat{ \boldsymbol{\mu } } } b_{ \hat{ \boldsymbol{\lambda } } \,,  \hat{ \boldsymbol{\mu } } } \, \hat{ \boldsymbol{\mu } } \,,
\eeqn
with $\hat{ \boldsymbol{\lambda } }$ being the irreps of the $\hat \gG_{  k }$, and $b_{ \hat{ \boldsymbol{ \lambda } } \,, \hat{ \boldsymbol{\mu } }}$ being the multiplicity of the irreps $\hat{ \boldsymbol{\mu } }$ of the $\oplus_{i} \hat \hG_{ i\,, k_i }$.
This leads to the equality between the conformal dimensions as follows
\beqn
%&& h_{ \hat {\boldsymbol{\lambda } }} + n = h_{\hat {\boldsymbol{\mu }} }\,, \quad{\rm or}~~ \frac{ ( \boldsymbol{\lambda } \,, \boldsymbol{\lambda } + 2 \boldsymbol{\rho } ) }{ 2(1+g^\lor ) } + n = \frac{ ( \boldsymbol{\mu } \,, \boldsymbol{\mu } + 2 \boldsymbol{\rho } ) }{ 2( x_e +p^\lor ) } \,, \non
&& h_{ \hat {\boldsymbol{\lambda } }} + n = h_{\hat {\boldsymbol{\mu }} }\,, \quad  h_{ \hat {\boldsymbol{\lambda } }}  = \frac{ C_2( \boldsymbol{\lambda } )_{\tt alg} }{ 2( 1+ g^\lor ) } \,, \quad  h_{ \hat {\boldsymbol{\mu } }} = \frac{ C_2(  \boldsymbol{\mu } )_{\tt alg} }{ 2( k_i + h_i^\lor ) } \,,
\eeqn
for the decompositions in Eqs.~\eqref{eqs:SU8irrep_decomp} with some grade $n$.
For the Abelian $\hat\uG(1)_{k_0 }$, there is no dual Coxeter number $g^\lor$ or the Weyl vector $\boldsymbol{\rho}$, hence the corresponding conformal dimension becomes
\beqn\label{eq:conformal_dim_Abelian}
h( \hat\uG(1))&=& \frac{ ( \boldsymbol{\lambda } \,, \boldsymbol{\lambda } )}{2 k_0 }= \frac{ \Xc_{0 }^2 }{2 k_0} ( \boldsymbol{\omega }_j  \,, \boldsymbol{\omega }_j   ) \Rightarrow \frac{ \Xc_{0\,,{\tt alg} }^2 }{4 k_{0\,, {\tt alg} } }  \,,
\eeqn
with the normalization of the fundamental weights in Eq.~\eqref{eq:fund-weight}.
At the ${\rm grade}=0$ and the level of $k_{0\,,{\tt alg}}=4$, we find that
\beqs\label{eqs:SU8irrep_conformalweight}
\beqn
&&  \rep{8} \hookrightarrow ( \rep{4} \,, \rep{1}\,,  - 1 ) \oplus  (\rep{1}\,, \rep{4}  \,,  +1 ) ~:~ h( \rep{ 8})= \frac{ 7}{16} \,, \non
&& h ( \rep{4} \,, \rep{1}\,,  -1 ) = h (\rep{1}\,, \rep{4}  \,,  +1 ) = \frac{15/8 }{1 +4} + \frac{( \pm 1)^2}{4 \times 4 } = \frac{7}{16} \,, \\[1mm]
%%%%%%%%%%%%%%%%%%%%%%%%%%%%%%%%%%%%%%%%%%%%%
&& \rep{28} \hookrightarrow ( \rep{6}\,, \rep{ 1} \,, - 2 ) \oplus ( \rep{1}\,, \rep{ 6} \,, + 2 ) \oplus  ( \rep{4}\,, \rep{4} \,,  0) ~:~ h( \rep{ 28})= \frac{ 3}{4 }  \,,\non
&& h ( \rep{6} \,, \rep{1}\,,  - 2 ) = h (\rep{1}\,, \rep{6}  \,,  + 2 ) = \frac{5/2 }{1+4 } + \frac{( \pm 2 )^2}{4 \times 4 } = \frac{ 3}{4 } \,, \non
&& h ( \rep{4}\,, \rep{4} \,,  0) = 2 \times \frac{15/8}{ 1+4 } = \frac{3}{4} \,,\\[1mm]
%%%%%%%%%%%%%%%%%%%%%%%%%%%%%%%%%%%%%%%%%%%%%
&& \rep{56} \hookrightarrow ( \rep{ 1}\,, \repb{4} \,, +3 ) \oplus  ( \repb{ 4}\,, \rep{1} \,, -3 ) \oplus ( \rep{ 4}\,, \rep{6} \,, + 1  )  \oplus ( \rep{ 6}\,, \rep{4} \,, - 1  ) ~:~  h( \rep{ 56})= \frac{ 15 }{16}  \,, \non
&& h ( \rep{ 1}\,, \repb{4} \,, + 3  ) = h ( \repb{ 4}\,, \rep{1} \,, - 3  )  = \frac{15/8}{1+4 } + \frac{ (\pm 3/4)^2 }{2 \times (1/2)} = \frac{15}{ 16} \,, \non
&& h ( \rep{ 4}\,, \rep{6} \,, + 1 ) = h ( \rep{ 6}\,, \rep{4} \,, - 1  ) =  \frac{15/8}{1+4 } +  \frac{5/2}{1+4 } + \frac{ (\pm 1/4)^2 }{2 \times (1/2)}= \frac{15}{ 16} \,, \\[1mm]
%%%%%%%%%%%%%%%%%%%%%%%%%%%%%%%%%%%%%%%%%%%%%
&& \rep{70} \hookrightarrow  ( \rep{1} \,, \rep{1 } \,, -4 ) \oplus ( \rep{1} \,, \rep{1 } \,, +4 ) \oplus  ( \rep{4} \,, \repb{4} \,, + 2 )  \oplus  ( \repb{4} \,, \rep{4} \,, - 2 )   \oplus ( \rep{6 } \,, \rep{6 } \,, 0 )   ~:~  h( \rep{ 70 })= 1 \,, \non
&& h (  \rep{1} \,, \rep{1 } \,, -4  ) = h ( \rep{1} \,, \rep{1 } \,, +4  ) = \frac{(\pm 4)^2}{4 \times 4 } = 1 \,, \quad h( \rep{6 } \,, \rep{6 } \,, 0 )   = 2\times \frac{5/2}{1+4} =1 \,, \non
&& h (  \rep{4} \,, \repb{4 } \,, + 2  ) = h ( \repb{4} \,, \rep{4 } \,, -  2 ) = 2 \times \frac{ 15/8}{1+4 } +  \frac{ (\pm 2)^2 }{ 4 \times 4 }= 1   \,,
\eeqn
\eeqs
with $C_2(\rep{4})_{\tt alg}= C_2(\repb{4})_{\tt alg}=15/4$ and $C_2( \rep{6})_{\tt alg}=5$.
With the conformal dimension defined in Eq.~\eqref{eq:conformal_dim_Abelian}, all conformal dimensions of the fields in Tab.~\ref{tab:U1TU1PQ} are found to match through their decompositions during the maximal symmetry breaking pattern.

\para
With the above results, we find a conformal embedding of $\widehat{ {\sG\uG}}(4)_1 \oplus \widehat{\sG\uG}(4)_1 \oplus \hat {\uG}(1)_{4\,,{\tt alg}} \subset \widehat{\sG\uG}(8)_1$ in the algebraic basis.
%To convert into the physical basis, we have to use the following normalization of the scalar products of the simple roots and the fundamental weights
%
%
%\beqn
%&& (  \boldsymbol{\alpha }_j  \,,  \boldsymbol{\alpha }_j  ) =1 \Rightarrow ( \boldsymbol{\omega }_j  \,,  \boldsymbol{\omega }_j  ) =1\,,
%\eeqn
%
%
%since the Lie algebra generators are normalized by $\Tr( T_{\rm fund}^A T_{\rm fund}^B)= \frac{1}{2} \delta^{AB}$ in the physical basis.
The conformal dimension in Eq.~\eqref{eq:conformal_dim_Abelian} in the physical basis becomes
\beqn\label{eq:conformal_dim_Abelian_phys}
h( \hat\uG(1))&=& \frac{ \Xc_{0\,,{\tt phys} }^2 }{4 k_{0\,, {\tt phys} } }  \,.
\eeqn
With the relation of $\Xc_{0\,,{\tt phys} } = \frac{1}{4} \Xc_{0\,,{\tt alg} }$, we find that
\beqn
&& k_{0\,,{\tt phys}} = \frac{1}{16 }k_{0\,,{\tt alg}} = \frac{1}{4} \,,
\eeqn
since the conformal dimension should be invariant under the basis conversion from Eq.~\eqref{eq:conformal_dim_Abelian} to Eq.~\eqref{eq:conformal_dim_Abelian_phys}.
Thus, we conclude a conformal embedding of the affine Lie algebra of $\widehat{ {\sG\uG}}(4)_{k_s = 1} \oplus \widehat{\sG\uG}(4)_{ k_W = 1 } \oplus \hat {\uG}(1)_{ k_{0\,,{\tt phys} }  = \frac{1}{4}  } \subset \widehat{\sG\uG}(8)_{ k_U = 1}$ in the physical basis.
To achieve the gauge coupling unification, the following relation at the scale of $v_U$ must be satisfied
\beqn\label{eq:441affine_unification}
&&  \alpha_{ U } (v_U) = \alpha_{4 s} (v_U) =  \alpha_{4W} (v_U) = \frac{1}{4} \alpha_{X_0}  (v_U) \,.
\eeqn
The RGE behaviors displayed in Fig.~\ref{fig:RGE_mini} are inconsistent with the above relation.
In particular, the gauge coupling of the $\alpha_{X_0}$ in the minimal ${\rm SU}(8)$ theory was too small to reach the condition in Eq.~\eqref{eq:441affine_unification}.

%###################################################################
\section{The RGEs of the SUSY $\widehat{\sG\uG}(8)_{ k_U = 1}$}
\label{section:SUSY_SU8}
%###################################################################

\para
Historically, an $\Nc=1$ SUSY extension to the ${\sG\uG}(5)$ theory was first considered in Ref.~\cite{Dimopoulos:1981zb}, where a minimal extension of a chiral superfield of $\repb{5_H}$ is necessary to cancel the anomaly of the $\rep{5_H}$ in the minimal ${\sG\uG}(5)$ theory when it was promoted to be supersymmetric.
Moreover, two chiral superfields of $( \rep{5_H} \,, \repb{5_H} )$ formulate the holomorphic Yukawa coupling terms of
\beqn
W_Y&=& \repb{5_F} \rep{10_F} \repb{5_H} + \rep{10_F} \rep{10_F} \rep{5_H} \,,
\eeqn
in the ${\sG\uG}(5)$ superpotential.

\para
When promoting all ${\rm SU}(8)$ fields in Tab.~\ref{tab:U1TU1PQ} to be chiral superfields, the super partners of the $\repb{8_H}_{\,,  \omega}  \oplus \repb{28_H}_{\,, \dot \omega}$ bring the total anomaly of
\beqn
&& \sum_\omega {\rm Anom}( \repb{8_H}_{\,,  \omega} ) + \sum_{\dot \omega} {\rm Anom} (\repb{28_H}_{\,, \dot \omega} )  = -24\,.
\eeqn
Naively, there may be two possible SUSY extensions~\cite{Chen:2024deo} with the additional chiral superfields in the spectrum to cancel the anomaly
\beqs\label{eqs:SUSY_SU8_chirals}
\beqn
&& \{ H \}_{ {\rm I}} =\rep{8_H}^{  \omega}  \oplus \rep{28_H}^{ \dot \omega} \,,  ~ \omega = ( 3\,, {\rm IV}\,, {\rm V}\,, {\rm VI}) \,, ~  \dot \omega = (\dot 1\,, \dot 2\,, \dot {\rm VII}\,, \dot {\rm VIII}\,, \dot {\rm IX} )   \,,\label{eq:SUSY_SU8_chiralsI} \\[1mm]
%%%%%%%%%%%%%%%%%%%%%%%%%%%%%%%%%%%%%%%%%%%%%
&& \{ H \}_{ {\rm II}} = \rep{ 36_H}^\rho  \,, ~ \rho = ( \odot \,, \ominus ) \,.\label{eq:SUSY_SU8_chiralsII}
\eeqn
\eeqs
Unfortunately, the SUSY extension in Eq.~\eqref{eq:SUSY_SU8_chiralsII} involve two rank-$2$ symmetric chiral superfields, which can only appear as the descendant states in the spectrum. 
It turns out the corresponding conformal dimension violates the unitarity constraint in Eq.~\eqref{eq:conformal_unitarity}, and this rules out the extension in Eq.~\eqref{eq:SUSY_SU8_chiralsII}.
One also has to guarantee that all Yukawa coupling terms should still arise from the holomorphic superpotential.
Three renormalizable Yukawa couplings in Eq.~\eqref{eq:Yukawa_SU8} are obviously holomorphic terms in the superpotential, while the non-renormalizable term therein can be modified into the following holomorphic term of
\beqn
&& W_Y \supset \frac{ c_4 }{ M_{\rm pl} } \rep{56_F}  \rep{56_F}  \rep{28_{H}}^{ \dot \omega }  \rep{63_{H}}   \,.
\eeqn
Two $d=5$ direct Yukawa coupling terms in Eq.~\eqref{eq:d5_direct} are also holomorphic terms in the superpotential.

\para
Below, we derive the corresponding RGEs according to the SWW symmetry breaking pattern of Eq.~\eqref{eq:SU8_SWW} for the SUSY extension in Eq.~\eqref{eq:SUSY_SU8_chiralsI}.
Generically, the two-loop RGE of a gauge coupling of $\alpha_\Upsilon$ is given by~\cite{Machacek:1983tz}
\beqn\label{eq:gauge_RGE}
&& \frac{d \alpha_\Upsilon ( \mu ) }{ d\log \mu} = \frac{ b_\Upsilon^{ (1) } }{2\pi } \alpha_\Upsilon^2 ( \mu ) +  \Big( \sum_{ \Upsilon^\prime } \frac{b_{  \Upsilon \Upsilon^\prime }^{ ( 2 ) } }{ 8 \pi^2  } \alpha_{\Upsilon^\prime } ( \mu )  \Big) \cdot  \alpha_\Upsilon^2 ( \mu ) \,.
\eeqn
We will always assume a set of SUSY RGEs between the $v_{441}\leq \mu\leq v_{U}$, and the corresponding one- and two-loop $\beta$ coefficients are
\beqs\label{eqs:SUSY_betas}
\beqn
{\rm non-Abelian}~&:&~ b_\Upsilon^{ (1 ) } = - 3 C_2( \gG_\Upsilon ) +  \sum_{ \Phi  } T(  \Rc^\Phi_\Upsilon )  \,, \non
&&~ b_{\Upsilon \Upsilon^\prime }^{ (2 ) } = - \frac{ 34}{3 }  C_2 ( \gG_\Upsilon )^2  + \sum_{ \Phi } \[ 6 C_2 ( \Rc^\Phi_{\Upsilon^\prime} )  +4 C_2 ( \gG_\Upsilon )   \] T( \Rc^\Phi_\Upsilon ) \,,  \\[1mm]
%%%%%%%%%%%%%%%%%%%%%%%%%%%%%%%%%%%%%%%%%%%%%
{\rm Abelian}~&:&~ b_\Upsilon^{ (1 ) } =   \sum_\Phi (  \Xc^\Phi_\Upsilon )^2  \,, \non
&&~ b_{\Upsilon \Upsilon^\prime }^{ (2 ) } =   6   \sum_{ \Phi \,, \Upsilon^\prime } (  \Xc^\Phi_{\Upsilon^\prime } )^2   \cdot (  \Xc^\Phi_\Upsilon )^2   \,.
\eeqn
\eeqs
Here, $C_2( \gG_\Upsilon )$ is the quadratic Casimir, $T(  \Rc^\Phi_\Upsilon ) $ is the trace invariant of the chiral superfield $\Phi$ with $\Rc^\Phi_\Upsilon \in \gG_\Upsilon$ and the physical normalization of $T(\Box )_{\tt phys}=1/2$, and $\Xc^\Phi_\Upsilon$ are the ${\uG}(1)$ charges at different stages.
The non-SUSY RGEs between the scales of $v_{\rm EW}\leq \mu\leq v_{441}$ are contributed by the one- and two-loop $\beta$ coefficients of
\beqs\label{eqs:nonSUSY_betas}
\beqn
{\rm non-Abelian}~&:&~ b_\Upsilon^{ (1 ) } = - \frac{11 }{ 3 } C_2( \gG_\Upsilon ) + \frac{2 }{3 } \sum_{ F  } T(  \Rc^F_\Upsilon ) +  \frac{1 }{3 } \sum_{ S   } T(  \Rc^S_\Upsilon ) \,,  \non
&&~ b_{\Upsilon \Upsilon^\prime }^{ (2 ) } = - \frac{ 34}{3 }  C_2 ( \gG_\Upsilon )^2  + \sum_{F  } \[ 2 \sum_{ \Upsilon^\prime } C_2 ( \Rc^F_{\Upsilon^\prime} )  + \frac{ 10}{ 3} C_2 ( \gG_\Upsilon )   \] T( \Rc^F_\Upsilon )\non
&& + \sum_S \[ 4 \sum_{ \Upsilon^\prime }  C_2 ( \Rc^S_{\Upsilon^\prime} ) + \frac{2 }{ 3 } C_2 ( \gG_\Upsilon )   \] T( \Rc^S_\Upsilon )\,, \\[1mm]
%%%%%%%%%%%%%%%%%%%%%%%%%%%%%%%%%%%%%%%%%%%%%
{\rm Abelian}~&:&~ b_\Upsilon^{ (1 ) } =  \frac{2 }{3 } \sum_F (  \Xc^F_\Upsilon )^2 +  \frac{1 }{3 } \sum_S ( \Xc^S_\Upsilon )^2\,,  \non
&& ~ b_{\Upsilon \Upsilon^\prime }^{ (2 ) } =   2    \sum_{ F\,, \Upsilon^\prime } (  \Xc^F_{\Upsilon^\prime } )^2   \cdot (  \Xc^F_\Upsilon )^2  + 4    \sum_{ S\,, \Upsilon^\prime }  (  \Xc^S_{\Upsilon^\prime } )^2  \cdot (  \Xc^S_\Upsilon )^2 \,.
%%%%%%%%%%%%%%%%%%%%%%%%%%%%%%%%%%%%%%%%%%%%%
%b_{ \Upsilon \Upsilon^\prime }^{ ( 2  ) }&=& - \frac{ 34}{3 }  C_2 ( \Gc_\Upsilon )^2 \alpha_\Upsilon - \sum_{F  } \[ 2 \sum_{ \Upsilon^\prime } C_2 ( \Rc^F_{\Upsilon^\prime} ) \alpha_{ \Upsilon^\prime } + \frac{ 10}{ 3} C_2 ( \Gc_\Upsilon ) \alpha_{ \Upsilon}  \] T( \Rc^F_\Upsilon )\non
%&& - \sum_S \[ 4 \sum_{ \Upsilon^\prime }  C_2 ( \Rc^S_{\Upsilon^\prime} ) \alpha_{ \Upsilon^\prime } + \frac{2 }{ 3 } C_2 ( \Gc_\Upsilon )\alpha_{ \Upsilon}  \] T( \Rc^S_\Upsilon ) \,,
\eeqn
\eeqs
When converting the SUSY RGEs to the non-SUSY RGEs at the $\mu=v_{441}$, a correction term~\cite{Antoniadis:1982vr,Langacker:1992rq,Martin:1993zk} of
\beqn
&& \Delta_\Upsilon = - \frac{ C_2 ( \gG_\Upsilon ) }{ 12 \pi }
\eeqn
should be accounted for the difference between the $\overline{\rm DR}$ scheme in the SUSY regime and the $\overline{\rm MS}$ scheme in the non-SUSY regime.

%\subsection{The RGEs of the SUSY $\widehat{\sG\uG}(8)_{ k_U = 1} $}

\para
Between the $v_{441}\leq \mu\leq v_{U}$, all chiral superfields besides of $\rep{63_H}$ in Tab.~\ref{tab:U1TU1PQ} are assumed to be massless.
Specifically, all chiral superfields in the second columns of Tabs.~\ref{tab:SU8_8barferm}, \ref{tab:SU8_28ferm}, and \ref{tab:SU8_56ferm} are massless, together with the following massless chiral superfields
\beqn
&& (  \repb{4} \,, \rep{1} \,, +\frac{1}{4} )_{\rep{H}\,,\omega }  \oplus ( \rep{1} \,, \repb{4} \,, -\frac{1}{4} )_{\rep{H}\,,\omega }  \subset  \repb{8}_{\rep{H}\,, \omega } \,,\non
&& (\rep{6 } \,,   \rep{1 } \,, +\frac{1}{2} )_{\rep{H}\,, \dot \omega } \oplus  ( \rep{1 } \,, \rep{6 } \,, -\frac{1}{2} )_{\rep{H}\,, \dot \omega } \oplus ( \repb{4 } \,, \repb{4 } \,, 0 )_{\rep{H}\,, \dot \omega }  \subset \repb{28}_{\rep{H}\,,\dot \omega } \,,\non
&&   ( \rep{4 } \,, \repb{4} \,, +\frac{1 }{ 2 } )_{\rep{H} } \oplus  (\repb{4 } \,, \rep{4} \,, -\frac{1 }{ 2}  )_{\rep{H} }  \oplus (\rep{6 } \,, \rep{6} \,,  0  )_{\rep{H} } \oplus (\rep{1}\,,\rep{1}\,,+1)_{\rep{H} } \oplus (\rep{1}\,,\rep{1}\,,-1)_{\rep{H} }\subset \rep{70_H} \,,\non
&&  (  \rep{4} \,, \rep{1} \,, -\frac{1}{4} )_{\rep{H} }^\omega \oplus ( \rep{1} \,, \rep{4} \,, +\frac{1}{4} )_{\rep{H} }^\omega  \subset  {\rep{8}_{\rep{H} }}^\omega  \,, \non
&& (\rep{6} \,, \rep{1 } \,, -\frac{1}{2} )_{\rep{H} }^{ \dot \omega } \oplus  ( \rep{1 } \,, \rep{6 } \,, +\frac{1}{2} )_{\rep{H} }^{\dot \omega } \oplus ( \rep{4} \,, \rep{4} \,, 0 )_{\rep{H} }^{ \dot \omega }  \subset {\rep{28}_{\rep{H}}}^{ \dot \omega} \,.
\eeqn
Correspondingly, we find that
\beqn
&&( b_{4s}^{(1)}\,, b_{4W}^{(1)}\,, b_{X_0}^{(1)}) =( 47, 47, 59) \,, \non
&&b_{\Gc_{441}}^{(2)}
= \left( \ba{ccc}
3181 /2&945/2& 30\\
945/2& 3181/2& 30\\
450& 450& 93\\
\ea \right)   \,,
\eeqn
according to Eqs.~\eqref{eqs:SUSY_betas}.

\para
Between the $v_{341}\leq \mu\leq v_{441}$, the spectrum contains massless fermions of
\beqn\label{eq:341_fermions}
&& \Big[ ( \repb{3}\,, \rep{1}\,, +\frac{1}{3} )_{ \mathbf{F}}^\Omega \oplus ( \rep{1}\,, \rep{1}\,, 0 )_{ \mathbf{F}}^\Omega \Big]  \oplus ( \rep{1}\,, \repb{4}\,, -\frac{1}{4} )_{ \mathbf{F}}^\Omega  \subset \repb{8_F}^\Omega \,, \non
&& \Omega = ( \omega \,, \dot \omega ) \,,\quad \omega = (3\,, {\rm V}\,, {\rm VI} ) \,,\quad \dot \omega = ( \dot 1\,, \dot 2\,, \dot {\rm VII} \,,\dot {\rm VIII} \,, \dot {\rm IX}) \,,\non
&& ( \rep{1}\,, \rep{1}\,, 0 )_{ \mathbf{F}}^{{\rm IV} } \subset \repb{8_F}^{ {\rm IV}} \,,\non
%%%%%%%%%%%%%%%%%%%%%%%%%%%%%%%%%%%%%%%%%%%%%
&&  ( \repb{3}\,, \rep{1}\,, -\frac{2}{3} )_{ \mathbf{F}}  \oplus ( \rep{1}\,, \rep{6}\,, +\frac{1}{2} )_{ \mathbf{F}}  \oplus  ( \rep{3}\,, \rep{4}\,, -\frac{1}{12} )_{ \mathbf{F}}  \subset \rep{28_F}\,, \non
%%%%%%%%%%%%%%%%%%%%%%%%%%%%%%%%%%%%%%%%%%%%%
&& ( \rep{1}\,, \repb{4}\,, +\frac{3}{4} )_{ \mathbf{F}} \oplus \Big[ ( \repb{3}\,, \rep{1}\,, -\frac{2}{3} )_{ \mathbf{F}}^\prime \oplus ( \rep{1}\,, \rep{1}\,, -1 )_{ \mathbf{F}} \Big] \oplus  \Big[ ( \rep{3}\,, \rep{6}\,, +\frac{1}{6} )_{ \mathbf{F}} \oplus ( \rep{1}\,, \rep{6}\,, +\frac{1}{2})_{ \mathbf{F}}^\prime \Big] \non
&\oplus& \Big[ ( \rep{3}\,, \rep{4}\,, -\frac{1}{12} )_{ \mathbf{F}}^\prime \oplus ( \repb{3}\,, \rep{4}\,, -\frac{5}{12})_{ \mathbf{F}} \Big]   \subset \rep{56_F}\,,
\eeqn
together with the massless Higgs fields of
\beqn
&& ( \rep{1} \,, \repb{4} \,, -\frac{1}{4} )_{\rep{H}\,, 3\,, {\rm V} \,, {\rm VI}}  \subset  \repb{8_H}_{ \,,3\,, {\rm V} \,, {\rm VI}} \,, \non
&&   ( \rep{1 } \,, \rep{6 } \,, -\frac{1}{2} )_{\rep{H}\,, \dot 1\,, \dot 2\,,  \dot {\rm VIII} } \subset \repb{28_H}_{ \,,\dot 1\,, \dot 2\,, \dot {\rm VIII}} \,, \non
&&  ( \rep{1} \,, \repb{4} \,, -\frac{1}{4} )_{\rep{H}\,, \dot 1 \,, \dot 2\,, \dot {\rm VII} \,, \dot {\rm IX} }  \subset ( \repb{4 } \,, \repb{4} \,, 0 )_{\rep{H}\,, \dot 1\,,  \dot 2\,, \dot {\rm VII} \,, \dot {\rm IX} } \subset \repb{28_H}_{ \,,\dot 1\,, \dot 2\,, \dot {\rm VII} \,, \dot {\rm IX} } \,,\non
&&  ( \rep{1 } \,, \repb{4} \,, +\frac{3 }{ 4} )_{\rep{H} }^{\prime  } \subset ( \rep{4 } \,, \repb{4} \,, +\frac{1 }{ 2 } )_{\rep{H} }  \subset \rep{70_H} \,, \non
&&  ( \rep{1} \,, \rep{4} \,, +\frac{1}{4} )_{\rep{H} }^{ \dot 1}  \subset ( \rep{4} \,, \rep{4} \,, 0 )_{\rep{H}  }^{ \dot 1} \subset \rep{28_H}^{ \dot 1}  \,.
\eeqn
Correspondingly, we find that
\beqn\label{eq:341_beta}
&&( b_{3c}^{(1)}\,, b_{4W}^{(1)}\,, b_{X_1}^{(1)}) =(-\frac{5}{3},-\frac{17}{6},+\frac{497}{36} ) \,, \non
&&b_{\Gc_{341}}^{(2)} =\left( \ba{ccc}
226/3&  75/2 &  97/36 \\
20&887 /6 &  95 /12 \\
194 /9 & 475 /4 & 4333 /216 \\
\ea \right)  \,,
\eeqn
according to Eqs.~\eqref{eqs:nonSUSY_betas}.

\para
Between the $v_{331}\leq \mu\leq v_{341}$, the spectrum contains massless fermions of
\beqn\label{eq:331_fermions}
&& \Big[ ( \repb{3}\,, \rep{1}\,, +\frac{1}{3} )_{ \mathbf{F}}^\Omega \oplus ( \rep{1}\,, \rep{1}\,, 0 )_{ \mathbf{F}}^\Omega \Big]  \oplus \Big[ ( \rep{1}\,, \repb{3}\,, -\frac{1}{3} )_{ \mathbf{F}}^\Omega \oplus ( \rep{1}\,, \rep{1}\,, 0 )_{ \mathbf{F}}^{\Omega^{\prime \prime }}  \Big]  \subset \repb{8_F}^\Omega \,, \non
&& \Omega = ( \omega \,, \dot \omega ) \,, \quad \omega = (3\,, {\rm VI})\,,  \quad \dot \omega = (\dot 1\,, \dot 2\,, \dot {\rm VIII}\,, \dot {\rm IX} )\,,\non
&& ( \rep{1}\,, \rep{1}\,, 0)_{ \mathbf{F}}^{{\rm IV} } \subset \repb{8_F}^{{\rm IV} } \,, \quad ( \rep{1}\,, \rep{1}\,, 0)_{ \mathbf{F}}^{{\rm V} \,, {\rm V}^{\prime\prime} } \subset \repb{8_F}^{{\rm V} } \,, \non
&&  ( \rep{1}\,, \rep{1}\,, 0)_{ \mathbf{F}}^{\dot {\rm VII} } \oplus ( \rep{1}\,, \rep{1}\,, 0)_{ \mathbf{F}}^{\dot {\rm VII}^{ \prime \prime} } \subset \repb{8_F}^{\dot {\rm VII} } \,,  \non
%%%%%%%%%%%%%%%%%%%%%%%%%%%%%%%%%%%%%%%%%%%%%
&&   ( \repb{3}\,, \rep{1}\,, -\frac{2}{3} )_{ \mathbf{F}}  \oplus  ( \rep{1}\,, \repb{3}\,, +\frac{2}{3} )_{ \mathbf{F}}  \oplus  ( \rep{3}\,, \rep{3}\,, 0 )_{ \mathbf{F}}    \subset \rep{28_F}\,, \non
%%%%%%%%%%%%%%%%%%%%%%%%%%%%%%%%%%%%%%%%%%%%%
&& ( \rep{1}\,, \repb{3}\,, +\frac{2}{3} )_{ \mathbf{F}}^\prime  \oplus  ( \repb{3}\,, \rep{1}\,, -\frac{2}{3} )_{ \mathbf{F}}^\prime  \oplus  \Big[ ( \rep{3}\,, \rep{3}\,, 0 )_{ \mathbf{F}}^\prime \oplus ( \rep{1}\,, \repb{3}\,, +\frac{2}{3})_{ \mathbf{F}}^{\prime \prime} \Big]  \non
&\oplus& \Big[ ( \rep{3}\,, \rep{3}\,, 0)_{ \mathbf{F}}^{\prime\prime}  \oplus ( \repb{3}\,, \rep{1}\,, -\frac{2}{3})_{ \mathbf{F}}^{\prime\prime\prime} \Big]   \subset \rep{56_F}\,.
\eeqn
together with the massless Higgs fields of
\beqn
&&  ( \rep{1} \,, \repb{3} \,, -\frac{1}{3} )_{\rep{H}\,, 3\,, {\rm VI} } \subset ( \rep{1} \,, \repb{4} \,, -\frac{1}{4} )_{\rep{H}\,, 3\,, {\rm VI} }  \subset  \repb{8_H}_{\,, 3\,, {\rm VI}} \,, \non
&&  ( \rep{1} \,, \repb{3} \,, -\frac{1}{3} )_{\rep{H}\,, \dot 2\,, \dot {\rm VIII} }^\prime   \subset ( \rep{1 } \,, \rep{6 } \,, -\frac{1}{2} )_{\rep{H}\,, \dot 2\,, \dot {\rm VIII} } \subset \repb{28_H}_{ \,, \dot 2\,, \dot {\rm VIII}} \,,\non
&&  ( \rep{1} \,, \repb{3} \,, -\frac{1}{3} )_{\rep{H}\,, \dot 2\,, \dot {\rm IX} }  \subset  ( \rep{1} \,, \repb{4} \,, -\frac{1}{4} )_{\rep{H}\,, \dot 2\,, \dot {\rm IX} }  \subset ( \repb{4 } \,, \repb{4} \,, 0 )_{\rep{H}\,, \dot 2\,, \dot {\rm IX} } \subset \repb{28_H}_{ \,,\dot 2\,, \dot {\rm IX}} \,, \non
&& ( \rep{1 } \,, \repb{3} \,, +\frac{2 }{ 3} )_{\rep{H} }^{\prime  \prime \prime } \subset   ( \rep{1 } \,, \repb{4} \,, +\frac{3 }{ 4} )_{\rep{H} }^{\prime  } \subset  ( \rep{4 } \,, \repb{4} \,, +\frac{1 }{ 2 } )_{\rep{H} }  \subset \rep{70_H} \,.
\eeqn
Correspondingly, we find that
\beqn\label{eq:331_beta}
&&( b_{3c}^{(1)} \,, b_{3W}^{(1)}\,, b_{X_2}^{(1)} ) =(-5,-\frac{23}{6},+\frac{82}{9}) \,, \non
&& b_{\Gc_{331}}^{(2)} =\left( \ba{ccc}
12& 12& 2 \\
12& 113 / 3 &38/9\\
16& 304 /9& 304/27\\
\ea \right)  \,,
\eeqn
according to Eqs.~\eqref{eqs:nonSUSY_betas}.

%%%%%%%%%%%%%%%%%%%%%%%%%%%%%%%%%%%%%%%%%%%%%%%%%%%%%%%
\begin{figure}[htb]
\centering
\includegraphics[height=6.5cm]{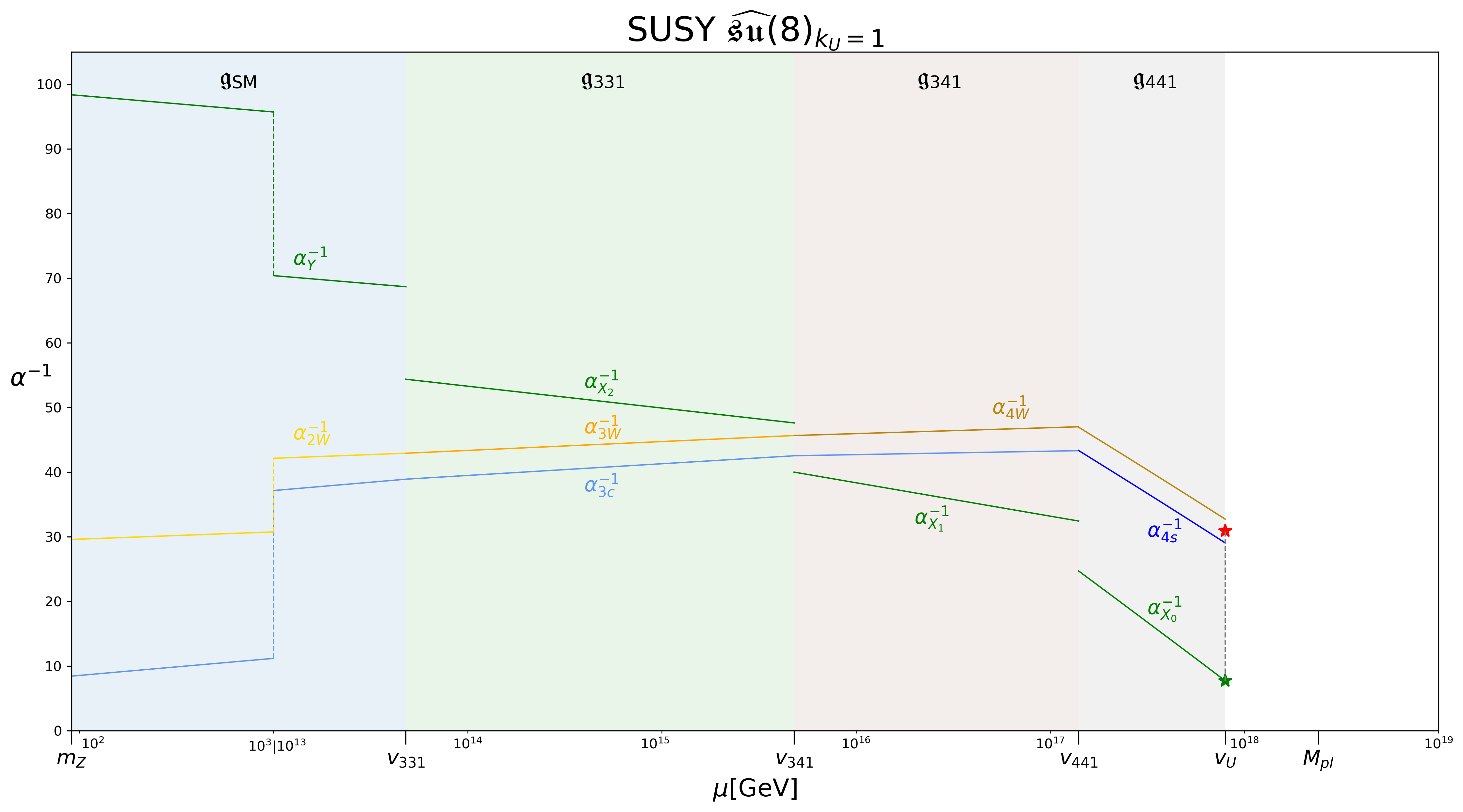}
\caption{The RGEs of the SUSY $\widehat{\sG\uG}(8)_{ k_U =1}$ setup.
The gauge couplings between $10^{3}\,{\rm GeV} \lesssim \mu \lesssim 10^{13}\,{\rm GeV}$ evolve according to the SM $\beta$ coefficients in Eq.~\eqref{eq:SM_beta} and are zoomed out in order to highlight the behaviors in three intermediate symmetry breaking scales.
}
\label{fig:RGE_TypeI}
\end{figure}
%%%%%%%%%%%%%%%%%%%%%%%%%%%%%%%%%%%%%%%%%%%%%%%%%%%%%%%

\para
Between the $v_{\rm EW}\leq \mu\leq v_{331}$, the massless fields include three-generational SM fermions together with one SM Higgs doublet of
\beqn\label{eq:SM_Higgs}
&& (\rep{1 } \,, \repb{2} \,, +\frac{1 }{ 2} )_{\rep{H} }^{\prime  \prime \prime } \subset ( \rep{1 } \,, \repb{3} \,, +\frac{2 }{ 3} )_{\rep{H} }^{\prime  \prime \prime } \subset   ( \rep{1 } \,, \repb{4} \,, +\frac{3 }{ 4} )_{\rep{H} }^{\prime  } \subset  ( \rep{4 } \,, \repb{4} \,, +\frac{1 }{ 2 } )_{\rep{H} }  \subset \rep{70_H} \,,
\eeqn
and we have the usual SM $\beta$ coefficients of
\beqn\label{eq:SM_beta}
&&( b_{3c}^{(1)}\,, b_{2W}^{(1)}\,, b_{Y}^{(1)} ) =(-7,-\frac{19}{6},+\frac{41}{6}) \,, \non
&& b_{\Gc_{\rm SM}}^{(2)}
=\left( \ba{ccc}
-26&  9/2 &11 /6 \\
12&35 /6&3 /2 \\
44/3 &9 /2&199 / 18 \\
\ea \right)  \,.
\eeqn
The corresponding RGEs of the SUSY $\widehat{ {\sG\uG}}(8)_{ k_U =1}$ extension are plotted in Fig.~\ref{fig:RGE_TypeI}, with three intermediate scales given in Eq.~\eqref{eq:SU8_SWW}.
The benchmark point is marked by $\star$ and reads
\beqn\label{eq:benchmark_TypeI}
&& c_{\rm HSW}\approx0.72\,, \quad v_U\approx8.0\times10^{17}\,\mathrm{GeV}\,,\non
&&  \alpha^{-1}_{4s}(v_U)= \alpha^{-1}_{4W}(v_U) \approx 30.9\,,\quad \alpha^{-1}_{X_0}(v_U) \approx7.71 \,.
\eeqn
A natural $\Oc(1)$ Wilsonian coefficient of $c_{\rm HSW}$ in Eq.~\eqref{eq:HSW_Op} is found to compensate for the small discrepancy between two non-Abelian gauge couplings.

\section{Summary}
\label{section:discussions}
%###################################################################

\para
Though the previous results in Ref.~\cite{Chen:2024cht} suggests encouraging outcome in addressing the SM quark/lepton masses and the CKM mixing patterns, the minimal setup based on the intermediate scales in Eq.~\eqref{eq:SU8_SWW} cannot achieve the gauge coupling unification~\cite{Chen:2024deo} in the context of the field theory.
In this paper, we describe the gauge coupling unification in the SUSY extensions of the affine $\widehat{ {\sG\uG}}(8)_{ k_U=1 }$ Lie algebra.
The SUSY version is obtained from the field contents of the minimal setup in Tab.~\ref{tab:U1TU1PQ}, with the additional chiral superfields in Eq.~\eqref{eq:SUSY_SU8_chiralsI} that (i) cancel the anomalies, and (ii) satisfy the unitarity constraint.
According to the RGEs plotted in Fig.~\ref{fig:RGE_TypeI}, the SUSY extension achieves the gauge coupling unification in Eq.~\eqref{eq:441affine_unification}, and the benchmark point was numerically presented in Eq.~\eqref{eq:benchmark_TypeI}.
The unification scales of $\sim \Oc(10^{18})\,{\rm GeV}$ are precious close to the Planck scale, which further suggests the intrinsic connections between the GUT and the underlying quantum gravity theory, such as the string theory.

%\para
%For the minimal Georgi-Glashow ${\rm SU}(5)$ theory, one has to consider the conformal embedding of the affine Lie algebra of $\widehat{ {\sG\uG}}(3)_{1 } \oplus \widehat{\sG\uG}(2)_{1 } \oplus \hat {\uG}(1)_{k_Y} \subset \widehat{\sG\uG}(5)_1$. The conformal dimensions can only match with $k_{Y\,,{\tt alg}}=15$ since
%
%
%\beqs\label{eqs:SU5irrep_conformalweight}
%\beqn
%&&  \repb{5} \hookrightarrow ( \repb{3 } \,, \rep{1}\,,  +2 ) \oplus  (\rep{1}\,, \repb{2}  \,,  -3 ) ~:~ h( \repb{ 5})= \frac{ 2}{ 5} \,, \non
%&& h ( \repb{3 } \,, \rep{1}\,,  +2 ) = \frac{4}{3} \times \frac{1}{1+3} + \frac{(+2)^2 }{4 \times 15}  = \frac{ 2}{ 5} \,, \non
%&& h (\rep{1}\,, \repb{2}  \,,  -3 ) = \frac{3}{4 } \times \frac{1}{1+2} + \frac{ (-3)^2 }{4 \times 15 } = \frac{ 2}{ 5} \,, \\[1mm]
%%%%%%%%%%%%%%%%%%%%%%%%%%%%%%%%%%%%%%%%%%%%%
%&& \rep{10} \hookrightarrow ( \rep{3}\,, \rep{ 2} \,, +1 ) \oplus ( \repb{3 }\,, \rep{ 1} \,, -4 ) \oplus  ( \rep{1}\,, \rep{1} \,,  +6) ~:~ h( \rep{ 10})= \frac{ 3}{5 }  \,,\non
%&& h ( \rep{6} \,, \rep{1}\,,  - 2 ) = h (\rep{1}\,, \rep{6}  \,,  + 2 ) = \frac{5/2 }{1+4 } + \frac{( \pm 2 )^2}{4 \times 4 } = \frac{ 3}{4 } \,, \non
%&& h ( \rep{4}\,, \rep{4} \,,  0) = 2 \times \frac{15/8}{ 1+4 } = \frac{3}{4}\,.
%\eeqn
%\eeqs
%
%
%Accordingly, we find that $k_{Y\,, {\tt phys}}=\frac{5}{12}$ in the physical basis.

\para
More generally, we prove the following conformal embedding of
\beqn\label{eq:SUN_k1_embedding}
&& \widehat{ {\sG\uG}}( n_s )_{k_s =1 } \oplus \widehat{\sG\uG}( n_W )_{k_W=1 } \oplus \hat {\uG}(1)_{k_{1\,,{\tt phys } } = \frac{1}{4 } } \subset \widehat{\sG\uG}(N)_{k_U = 1 } \,, \quad n_s + n_W = N \,,
\eeqn
in the physical basis.
The ${\rm U}(1)_1$ charges of the fundamental representation in the physical basis are normalized as~\footnote{Notice that, one has to evaluate the RGE of the ${\uG}(1)_1$ gauge coupling for the unification in any unified theory. For example, the  ${ \uG}(1)_1$ gauge coupling in the Georgi-Glashow theory of ${\sG\uG}(5) \hookrightarrow \gG_{\rm SM}$ is normalized as $g_1=\sqrt{ \frac{ 5}{ 3} } g_Y$. In the symmetry breaking patterns of ${\sG\uG}(8) \hookrightarrow {\sG\uG}(4) \oplus {\sG\uG}(4) \oplus {\uG}(1) $ or ${\sG\uG}(9) \hookrightarrow  {\sG\uG}(6) \oplus  {\sG\uG}(3) \oplus {\uG}(1) $, no additional normalizations are needed.}
\beqn
&& \rep{N} = ( \rep{n_s} \,, \rep{1} \,,  -  \frac{n_W}{ \sqrt{ 2 n_s n_W N } } ) \oplus ( \rep{1} \,, \rep{n_W } \,,  +    \frac{n_s}{ \sqrt{ 2 n_s n_W N } } ) \,.
\eeqn
Thus, the equality between the conformal dimensions leads to
\beqn
&&  \frac{ N-1 }{2N }   = \frac{ n_s -1 }{ 2 n_s }  + \frac{1}{ 4 k_{ 1 \,,{\tt phys} }} \Big(  -  \frac{n_W}{ \sqrt{ 2 n_s n_W N  } }   \Big)^2 =  \frac{ n_W -1 }{ 2 n_W }  + \frac{1}{ 4 k_{ 1 \,,{\tt phys} }} \Big(  +  \frac{n_s }{ \sqrt{ 2 n_s n_W N  } }   \Big)^2 \non
&\Rightarrow& k_{ 1 \,,{\tt phys} } = \frac{1}{4} \,.
\eeqn
This means the conformal embedding in Eq.~\eqref{eq:SUN_k1_embedding} always leads to an affine level of $k_{ 1\,,{\tt phys} } = \frac{1}{4}$ in the physical basis, regardless of the specific symmetry breaking pattern.
This means the gauge coupling unification is generalized from the relation in Eq.~\eqref{eq:441affine_unification} to the following
\beqn\label{eq:affine_unification}
&&  \alpha_{ U } (v_U) = \alpha_{ s} (v_U) =  \alpha_{ W} (v_U) = \frac{1 }{ 4} \alpha_{ 1}  (v_U) \,,
\eeqn
regardless of the specific symmetry breaking patterns in the theory based any $\widehat{ \sG \uG}( N)_{k_U=1}$ affine Lie algebra.

\para
For the SUSY $\widehat{ {\sG\uG}}(8)_{ k_U =1 }$ theory, there may be non-maximally symmetry breaking patterns of
\beqs
\beqn
&& {\sG\uG}(8) \xhookrightarrow{ \langle  \rep{63_H}\rangle } \gG_{531}/\gG_{351} \,,  \non
&& \gG_{531}  \equiv {\sG\uG}(5)_{s} \oplus {\sG\uG}(3)_W \oplus  {\uG}(1)_{X_0 } \,, \quad  \gG_{351}  \equiv {\sG\uG}(3)_{s} \oplus {\sG\uG}(5)_W \oplus {\uG}(1)_{X_0 }  \,, \non
{\rm embedding}~&:&~ \widehat{ {\sG\uG}}(5)_{k_{s/W} =1 } \oplus \widehat{\sG\uG}(3)_{k_{W/s} =1 } \oplus \hat {\uG}(1)_{k_{1\,,{\tt phys}} = \frac{1}{4} } \subset \widehat{\sG\uG}(8)_{ k_U = 1} \,,\\[1mm]
%%%%%%%%%%%%%%%%%%%%%%%%%%%%%%%%%%%%%%%%%%%%%
&& {\sG\uG}(8)   \xhookrightarrow{ \langle  \rep{63_H}\rangle }\gG_{621} \,,\quad \gG_{621}  \equiv {\sG\uG}(6)_{s} \oplus {\sG\uG}(2)_W \oplus {\uG} (1)_{X_0 } \,, \non
{\rm embedding}~&:&~ \widehat{ {\sG\uG}}(6)_{k_s } \oplus \widehat{\sG\uG}(2)_{k_W} \oplus \hat {\uG}(1)_{k_{1\,,{\tt phys}} = \frac{1  }{ 4 } } \subset \widehat{\sG\uG}(8)_{ k_U = 1 } \,,
\eeqn
\eeqs
according to Witten~\cite{Witten:1981nf}.
Along both symmetry breaking patterns, the conformal invariance in Eq.~\eqref{eq:cc_relation} fixes the levels of both subalgebras to be
\beqn
{\sG\uG}(8) \hookrightarrow \gG_{531}/\gG_{351}~&:&~ \frac{24\, k_s }{k_s + 5 } + \frac{ 8\, k_W }{k_W + 3}  + 1 = 7 \Rightarrow ( k_s \,, k_W ) = (1\,, 1 ) \,, \non
{\sG\uG}(8) \hookrightarrow \gG_{621}~&:&~ \frac{35\, k_s }{k_s + 6 } + \frac{ 3\, k_W }{k_W + 2}  + 1 = 7 \Rightarrow ( k_s \,, k_W ) = (1\,, 1 ) \,.
\eeqn
Therefore, these two non-maximally symmetry breaking patterns can only achieve the gauge coupling unification in the context of the conformal embedding when the relation in Eq.~\eqref{eq:affine_unification} holds.

\begin{figure}[htb]
\centering
\includegraphics[height=6cm]{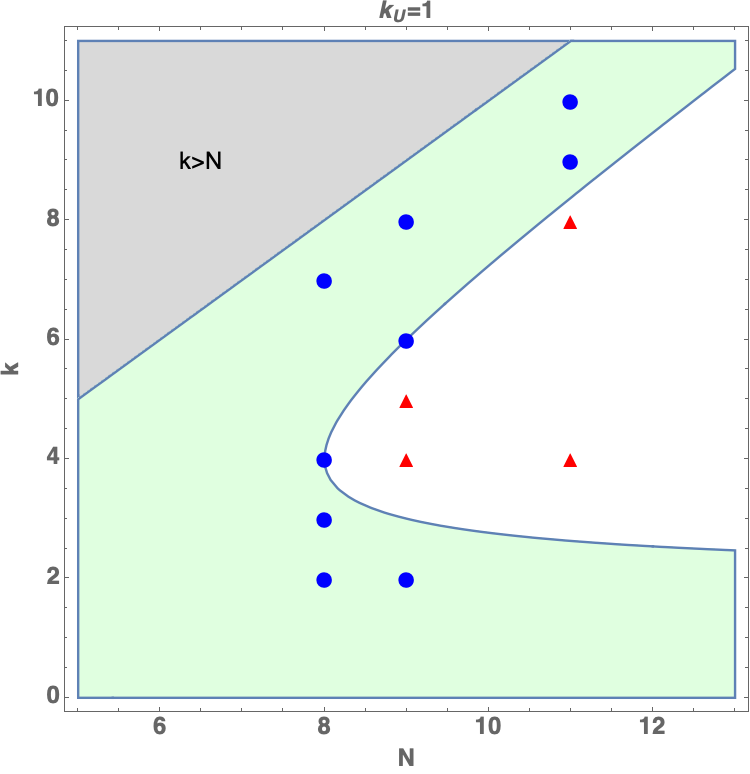}
\caption{The unitarity allowed region (green shaded) to the rank-$k$ anti-symmetric irreps of the $\widehat{ \sG \uG}(N)_{k_U =1}$ theories.
The blue circles and red triangles represent the allowed and the excluded irreps, respectively.
The gray shaded region with $k>N$ is not considered.
}
\label{fig:Ak_Unitarity}
\end{figure}
%%%%%%%%%%%%%%%%%%%%%%%%%%%%%%%%%%%%%%%%%%%%%%%%%%%%%%%

\para
Finally, we comment on the conformal embedding for the non-minimal flavor-unified theories beyond the $\widehat{ \sG \uG}(8)_{ k_U =1}$.
Given the RGE behaviors in the minimal ${\sG\uG}(8)$ theory, it is reasonable to conjecture that the similar behaviors can also exist when one extends the flavor-unified theories beyond the affine Lie algebra of $\widehat{ \sG \uG}(8)$.
Here, we list the matter contents in two possible non-minimal (and non-SUSY) extensions of
\beqs\label{eqs:nonmini_FlavorGUT}
\beqn
{\sG\uG}(9)~&:&~ \{ f_L \}_{ {\sG\uG}(9)}^{n_g=3} = \Big[  \repb{9_F}^\omega \oplus \rep{36_F} \Big] \bigoplus \Big[  \repb{9_F}^{ \dot \omega } \oplus \rep{126_F} \Big] \,, \non
&&   \{ H \}_{ {\sG\uG}(9)}^{n_g=3} =  \repb{9_H}_{\,, \omega }  \oplus \repb{84_H}_{\,, \dot \omega }   \oplus \repb{126_H} \oplus  \rep{80_H}  \,, \non
&&  \omega = ( 3\,, {\rm IV}\,, {\rm V}\,, {\rm VI}\,, {\rm VII}) \,, ~  \dot \omega = (\dot 1\,, \dot 2 \,, \dot {\rm VIII}\,, \dot {\rm IX} \,, \dot {\rm X} ) \,,\\[1mm]
{\sG\uG}(11)~&:&~ \{ f_L \}_{ {\sG\uG}(11)}^{n_g=3} = \rep{330_F} \oplus \repb{ 165_F} \oplus \repb{ 55_F} \oplus  \repb{11_F} \,,
\eeqn
\eeqs
which were previously proposed in Refs.~\cite{Chen:2023qxi} and \cite{Georgi:1979md}. 
Both of them involve higher-rank anti-symmetric irreps and lead to three-generational SM fermions mathematically.
In Fig.~\ref{fig:Ak_Unitarity}, we plot the unitarity constrained regions~\eqref{eq:conformal_dim_nonAbelian} to the rank-$k$ anti-symmetric irreps $A_k$ from the $\widehat{ \sG \uG}( 8)_{k_U =1}$~\eqref{eq:SU8_3gen_fermions} \eqref{eq:Yukawa_SU8}, $\widehat{ \sG \uG}(9)_{k_U =1}$, and $\widehat{ \sG \uG}( 11)_{k_U =1}$ theories~\eqref{eqs:nonmini_FlavorGUT}.
Apparently, both the $\widehat{\sG \uG}(9)_{ k_U =1}$ and the $\widehat{\sG\uG}(11)_{ k_U =1}$ extensions contain the irreps excluded by the unitarity constraints, as compared to the allowed $\widehat{\sG\uG}(8)_{k_U =1}$ theory.
In other words, the gauge coupling unifications in such non-minimal extensions cannot be achieved with the relation of \eqref{eq:affine_unification} through the conformal embedding.

%###################################################################
\section*{Acknowledgements}
%###################################################################
%
%
\para
We would like to thank Kaiwen Sun, Yuan Sun, Yinan Wang, and Wenbin Yan for very enlightening discussions and communications.
N.C. would like to thank South China Normal University, University of Science and Technology of China, Hangzhou Institute for Advanced Study, and Liaoning Normal University for hospitality when preparing this work.
N.C. is partially supported by the National Natural Science Foundation of China (under Grants No. 12035008 and No. 12275140) and Nankai University.

\appendix

%###################################################################
\section{Some results of the affine Lie algebra}
\label{section:affine}
%###################################################################

\para
In this appendix, we review some necessary results of the Lie algebra and the affine Lie algebra.
We follow the conventions in the textbook of the conformal field theory~\cite{DiFrancesco:1997nk} closely.
All results described in this section are given in the algebraic basis.

%%%%%%%%%%%%%%%%%%%%%%%%%%%%
\subsection{The Lie algebra: Cartan-Weyl basis, simple roots, and the highest weight}
%%%%%%%%%%%%%%%%%%%%%%%%%%%%

\para
The Lie algebra of $\gG$ is defined according to the commutation relations among ${\rm dim}(\gG)$ generators as follows
\beqn\label{eq:Lie_commutator}
&& \[  T^A \,, T^B \] = i f^{ABC } T^C\,, \quad A\,,B\,,C = 1\,,...\,, {\rm dim}(\gG) \,,
\eeqn
where $f^{ABC}$ are known as the structure constants.
$\gG$ has a maximal set of commuting generators to form the Cartan subalgebra, and we denote them by
\beqn\label{eq:Lie_Cartan}
&& \{ \Hc^i \} \,, \quad i = 1\,, ... \,, r\,, \quad r={\rm rank}(\gG) \,, \quad \[ \Hc^i \,,  \Hc^j \] =0 \,.
\eeqn
All other ${\rm dim}(\gG)-r$ generators of $E^{ \boldsymbol{\alpha } }$ are the ladder operators, and their commutators with the Cartan generators lead to the roots $\boldsymbol{\alpha }$ of the $\gG$ as follows
\beqn\label{eq:Lie_Ladders_Cartan}
&& \[ \Hc^i \,, E^{ \boldsymbol{\alpha }} \] = \alpha^i E^{\boldsymbol{\alpha } } \,,\quad \boldsymbol{\alpha}  \equiv ( \alpha^1\,, ... \,, \alpha^r ) \,.
\eeqn
Since for any root $\boldsymbol{\alpha}$, the $- \boldsymbol{\alpha}$ is also a root, one thus partitions the set of all roots $\{ \boldsymbol{\alpha} \}$ in Eq.~\eqref{eq:Lie_Ladders_Cartan} into the set of positive root and negative roots as
\beqn
\Delta&\equiv& \Delta_+ \oplus \Delta_- \,,\quad {\rm dim} \Delta_+ = {\rm dim} \Delta_- = \hf \Big( {\rm dim}(\gG) - r  \Big) \,.
\eeqn
The mutual commutators between the raising operators of $E^{ \boldsymbol{\alpha }}$ and lowering operators of $E^{ -\boldsymbol{\alpha }}$ are
\beqn\label{eq:Lie_Ladder_Ladder}
&& \[  E^{ \boldsymbol{\alpha }} \,, E^{ - \boldsymbol{\alpha }}  \] = \frac{2}{ ( \boldsymbol{\alpha } \,,  \boldsymbol{\alpha })  } \boldsymbol{\alpha } \cdot \boldsymbol{\Hc } \,.
\eeqn
All roots can be generally written as linear combinations of other roots, and the $r$ linearly independent set of roots are known as the simple roots of $\boldsymbol{\alpha}_i$ (with $i=1\,,...\,,r$).
The simple coroots are defined according to the simple roots as
\beqn\label{eq:coroot}
\boldsymbol{\alpha}_i^\lor &\equiv& \frac{2 \boldsymbol{\alpha}_i }{ ( \boldsymbol{\alpha}_i \,, \boldsymbol{\alpha}_i )} \,.
\eeqn
%
%
%\NingC{[In some monopole literature, the definition differs by a factor $2$. We also have to clarify the definition of the scalar product.]}
For the $\sG \uG(N)$ Lie algebras, the scalar products between the simple roots are normalized as $( \boldsymbol{\alpha}_j \,, \boldsymbol{\alpha}_j )=2$, hence one has $\boldsymbol{\alpha}_i^\lor = \boldsymbol{\alpha}_i$ and $( \boldsymbol{\alpha}_j^\lor \,, \boldsymbol{\alpha}_j^\lor )=2$ as well.
The scalar products between the simple roots and the coroots define the $\sG \uG(N)$ Cartan matrix
\beqn\label{eq:Cartan_matrix}
&& A_{i\, j}  \equiv ( \boldsymbol{\alpha}_i \,, \boldsymbol{\alpha}_j^\lor ) \,, \non
&& A_{i \, i} =2 \,, \quad A_{i\, i\pm 1} ( 1< i < N-1 ) = A_{1\, 2 } = A_{ N-1 \, N-2 } =-1 \,.
\eeqn
Among all roots in the $\Delta$, there is a unique {\it highest root} $\boldsymbol{ \theta } $, which can be expressed as the summation of all simple roots or simple coroots 
\beqn\label{eq:highest_root}
 \boldsymbol{ \theta } &=& \sum_{i=1}^{ N-1} a_i \boldsymbol{\alpha}_i =  \sum_{i=1}^{ N-1 } a_i^\lor \boldsymbol{ \alpha } _i^\lor \,, \quad a_i \,, a_i^\lor \in \mathbb{Z}\,.
\eeqn
The $a_i$ and $a_i^\lor$ are the {\it marks} and the {\it comarks} and one has $a_i = a_i^\lor =1$ for the $\sG\uG(N)$ Lie algebra.
The highest root $\boldsymbol{ \theta } $ for the $\sG\uG(N)$ is given by
\beqn
\boldsymbol{ \theta } &=& ( 1\,, \underbrace{ 0 \,, ... \,, 0 }_{ N-3 } \,, 1)  \,.
\eeqn
The {\it Coxeter number} $g$ and the {\it dual Coxeter number} $g^\lor$ can be defined as
\beqs
\beqn
&& g \equiv \sum_{i=1}^{ N-1 } a_i + 1\,, \label{eq:Coxeter} \\[1mm]
&& g^\lor \equiv \sum_{i=1}^{ N-1 } a_i^\lor + 1\,.\label{eq:dualCoxeter}
\eeqn
\eeqs
The dual Coxeter number reads $g^\lor=N$ for the $\sG\uG(N)$ Lie algebra, while the Abelian $\uG(1)$ Lie algebra has no dual Coxeter number.

\para
The fundamental weights of $\boldsymbol{ \omega }_i$ are defined from the simple coroots as follows
\beqn\label{eq:fund-weight}
&& ( \boldsymbol{ \omega }_i \,, \boldsymbol{ \alpha } _j^\lor ) = \delta_{ij} \,.
\eeqn
The highest-weight states, denoted as $ \boldsymbol{ \lambda } $, are defined by
\beqn\label{eq:highest-weight}
&& E^{ \boldsymbol{ \alpha } } | \boldsymbol{ \lambda } \rangle =0 \,,\quad \forall~~ \boldsymbol{ \alpha } \in \Delta_+ \,,
\eeqn
and are expressed in terms of the Dynkin labels as follows
\beqn\label{eq:Dynkin-label}
&&  \boldsymbol{ \lambda }  = \sum_{i=1}^r \lambda_i \boldsymbol{ \omega }_i  \,, \quad \boldsymbol{ \lambda }  = ( \lambda_1\,, ... \,, \lambda_{ N-1 }  ) =   \{ \ell_1\,, ... \,, \ell_{N-1}  \} \,, \quad \lambda_i \,, \ell_i \in \mathbb{Z} \,,
\eeqn
with
\beqn
&&\ell_i = \lambda_i + ... + \lambda_{N-1}
\eeqn
representing the number of boxes in the $i^{\rm th}$ row of a corresponding Young tableau.
For instance, the Dynkin labels for the rank-$k$ anti-symmetric, the rank-$k$ symmetric, and the adjoint representations are expressed as follows
\beqs
\beqn
&&  \boldsymbol{ \lambda }  ( A_k)= ( \underbrace{ 0\,, ... \,,0 }_{k-1} \,,1\,, \underbrace{ 0 \,, ... \,, 0}_{ N-k - 1 } ) \,,\label{eq:Dynkin-Ak} \\[1mm]
&&  \boldsymbol{ \lambda }  ( S_k)= ( k\,, \underbrace{ 0\,, ... \,,0 }_{ N -2}  ) \,,\label{eq:Dynkin-Sk} \\[1mm]
&&  \boldsymbol{ \lambda }  ( {\rm adj} )=  ( 1\,, \underbrace{ 0 \,, ... \,, 0}_{ N-3 } \,, 1 ) \,.\label{eq:Dynkin-adj}
\eeqn
\eeqs
For any irrep, all weights can be constructed by subtracting the simple roots from the corresponding highest-weight state, and the total numbers of the weights through this procedure correspond to the dimension of the irrep.
The scalar products between the $\sG\uG(N)$ fundamental weights are expressed as the quadratic form matrix of
\beqn\label{eq:quadratic_form}
F_{ij}&=& (  \boldsymbol{ \omega }_i \,, \boldsymbol{ \omega }_j  ) = \hf ( \boldsymbol{ \alpha }_j \,, \boldsymbol{ \alpha }_j ) ( A^{-1} )_{ij} = \frac{1}{N} {\rm min} ( i \,, j ) \cdot ( N -{\rm max} (i\,, j ) )  \non
&=& \frac{1}{ N } \left( \ba{cccccc}
N-1  & N-2   & N-3  &  ... & 2 & 1 \\
 N-2 &  2(N-2) & 2(N-3)  & ...  &  4 &  2 \\
N-3  & 2(N-3)  & 3(N-3)  & ...  &  6 & 3  \\
.  & .  & .  & ...  &  . &  . \\
 2 & 4  & 6  & ...  & 2( N-2 )  & N-2  \\
1  & 2  & 3  & ...  &  N-2 &  N-1  \\
\ea \right)  \,,
\eeqn
where $(A^{-1})_{ij}$ is the inverse of the Cartan matrix~\eqref{eq:Cartan_matrix}, and we have used the normalization of $( \boldsymbol{ \alpha }_k \,, \boldsymbol{ \alpha }_k )=2$.
The scalar products between two highest-weight states are thus given by
\beqn\label{eq:scalar_prod_weights}
&& ( \boldsymbol{ \lambda }^{ (1) } \,,   \boldsymbol{ \lambda }^{ ( 2 ) }  ) = \sum_{i \,, j }  \lambda_i^{ ( 1 ) }  \lambda_j^{ ( 2 ) } (  \boldsymbol{ \omega }_i \,, \boldsymbol{ \omega }_j ) = \sum_{i \,, j }  \lambda_i^{ ( 1 ) }  \lambda_j^{ ( 2 ) } F_{ij} \,.
\eeqn
A special type of the weight is the Weyl vector, which is given by the summation over all fundamental weights, or half of the summation over all positive roots as follows
\beqn\label{eq:Weyl_vec}
\boldsymbol{ \rho } &\equiv& \sum_{i=1}^{ N-1} \boldsymbol{ \omega }_i = \hf \sum_{\boldsymbol{ \alpha }  \in \Delta_+ } \boldsymbol{ \alpha }  = ( \underbrace{1\,, ... \,, 1}_{ N-1 } ) \,.
\eeqn
The Abelian $\uG(1)$ Lie algebra has no Weyl vector.

\para
The quadratic Casimir operator of $\Qc$ is defined by
\beqn
\Qc&\equiv& \sum_i \Hc^i  \Hc^i + \sum_{ \boldsymbol{\alpha} \in \Delta_+ } ( E^{ \boldsymbol{\alpha} } E^{ - \boldsymbol{\alpha}  } + E^{ - \boldsymbol{\alpha} } E^{  \boldsymbol{\alpha}  } )\,.
\eeqn
in the Cartan-Weyl basis.
Since it commutes with all generators, it is mostly convenient to evaluate the eigenvalue of $C_2(  \boldsymbol{\lambda } )_{\tt alg}$ on the highest-weight state of $| \boldsymbol{\lambda } \rangle$, which reads~\footnote{The eigenvalue of the quadratic Casimir operator differs by a factor $2$ from the conventions in the particle physics.}
\beqn
&&  \sum_{i=1}^r \Hc^i  \Hc^i | \boldsymbol{\lambda } \rangle = (  \boldsymbol{\lambda } \,,  \boldsymbol{\lambda } ) | \boldsymbol{\lambda } \rangle \,, \non
&& \sum_{\boldsymbol{ \alpha }  \in \Delta_+ } ( E^{ \boldsymbol{\alpha} } E^{ - \boldsymbol{\alpha}  } + E^{ - \boldsymbol{\alpha} } E^{  \boldsymbol{\alpha}  } )  | \boldsymbol{\lambda } \rangle = \sum_{\boldsymbol{ \alpha }  \in \Delta_+ }  \[ E^{ \boldsymbol{\alpha} }   \,, E^{ - \boldsymbol{\alpha} }  \] | \boldsymbol{\lambda } \rangle = \sum_{\boldsymbol{ \alpha }  \in \Delta_+ }  ( {\boldsymbol{ \alpha } } \,,  {\boldsymbol{ \lambda }} ) | \boldsymbol{\lambda }  \rangle =  ( 2 {\boldsymbol{ \rho } } \,,  {\boldsymbol{ \lambda }} ) | \boldsymbol{\lambda }  \rangle \,, \non
&\Rightarrow& C_2(  \boldsymbol{\lambda } )_{\tt alg} = ( {\boldsymbol{ \lambda }} + 2 {\boldsymbol{ \rho } } \,,  {\boldsymbol{ \lambda }} ) \,,
\eeqn
where we have used the commutator in Eq.~\eqref{eq:Lie_Ladder_Ladder} and the definition of the Weyl vector in Eq.~\eqref{eq:Weyl_vec}.
For the rank-$k$ anti-symmetric representation in Eq.~\eqref{eq:Dynkin-Ak}, one finds that
\beqn\label{eq:Casimir-Ak}
C_2(   A_k  )_{\tt alg} &=&  ( \underbrace{ 0\,, ... \,,0 }_{k-1} \,,1\,, \underbrace{ 0 \,, ... \,, 0}_{ r-k } )  \cdot ( \underbrace{ 2\,, ... \,,2 }_{k-1} \,,3 \,, \underbrace{ 2 \,, ... \,, 2 }_{ r-k } ) \non
&=&  2  \frac{ N-k }{N} \sum_{j=1}^{k-1} j + 2 \frac{k }{ N} \sum_{j = k+1}^{N-1} ( N-j ) + 3 \frac{k (N-k ) }{N } =  \frac{ N+1 }{ N } k (N-k)   \,.
%&=&  \frac{N-k}{N} k ( k-1 ) + \frac{k }{ N} (N-k ) ( N-k -1 ) + \frac{3 k (N-k) }{N}  \non
%&=& \frac{ N+1 }{ N } k (N-k)\,.
\eeqn
For the rank-$k$ symmetric representation in Eq.~\eqref{eq:Dynkin-Sk}, one finds that
\beqn\label{eq:Casimir-Sk}
C_2(   S_k  )_{\tt alg} &=&  ( k \,, \underbrace{ 0\,, ... \,,0 }_{ N -2}  )  \cdot ( k+2 \,, \underbrace{ 2\,, ... \,,2 }_{ N -2}  )  \non
&=&  \frac{ N - 1 }{ N } k (k + 2 )  + 2 k \sum_{j=2}^{N-1}  \frac{ N-j }{N }   \non
&=& \frac{ N-1 }{ N } k (N+k)\,.
\eeqn
For the adjoint representation with the highest-weight state being the highest root of ${\boldsymbol{ \lambda }} = {\boldsymbol{ \theta }}$, one finds the well-known result of
\beqn\label{eq:Casimir-adj}
C_2(  \boldsymbol{\theta } )_{\tt alg} &=&    ( {\boldsymbol{ \theta }} + 2 {\boldsymbol{ \rho } } \,,  {\boldsymbol{ \theta }} ) = 2 + 2 ( {\boldsymbol{ \rho } }  \,, {\boldsymbol{ \theta } } ) \non
&=& 2 +2  \sum_{i\,, j=1}^{ N-1} a_i^\lor ( {\boldsymbol{\omega}}_j \,,  {\boldsymbol{\alpha}}_i^\lor ) = 2 + 2 \sum_{i=1}^{ N-1} a_i^\lor = 2 g^\lor \,,
\eeqn
by using the definition~\eqref{eq:dualCoxeter}.

%%%%%%%%%%%%%%%%%%%%%%%%%%%%
\subsection{The affine Lie algebra: Cartan-Weyl basis, simple roots, and the highest weight}
%%%%%%%%%%%%%%%%%%%%%%%%%%%%

\para
Next, we generalize the Lie algebra $\gG$ into the \textit{loop algebra} of $\tilde \gG$ as
\beqn
&& \tilde \gG = \gG \otimes {\mathbb C} \[ t \,, t^{-1} \] \,,
\eeqn
such that the commutation relations in Eq.~\eqref{eq:Lie_commutator} become
\beqn\label{eq:loop_commutator}
&& \[ {T^A}_m \,, {T^B}_n \] = i f^{ABC } {T^C}_{m+n}   \,,
\eeqn
with ${T^A}_m \equiv T^A \otimes t^m$.
A central extension to the commutator is given by
\beqn\label{eq:central_commutator}
&& \[  {T^A}_m \,, {T^B}_n \] = i f^{ABC } {T^C}_{m+n} + \hat k n \delta^{ AB} \delta_{ n + m \,,0} \,,
\eeqn
with
\beqn
&& \[  {T^A}_m \,, \hat k \] =0 \,.
\eeqn
In other words, the central extension only exists when $n+m=0$.
The affine Lie algebra is defined by
\beqn
\hat \gG &\equiv& \tilde \gG \oplus {\mathbb C} \hat k \oplus   {\mathbb C}  L_0 \,.
\eeqn
The $L_0$ is known as the grading operator of
\beqn\label{eq:grading_Op}
&& L_0 \equiv - t \frac{d }{ dt } \,,
\eeqn
and its action on the generators gives arise to the {\it grade} $n$ as below
\beqn
&& [  L_0 \,,  {T^A}_n ] = - n {T^A}_n \,.
\eeqn
The maximal Cartan subalgebra of the affine Lie algebra $\hat \gG$ is given by
\beqn\label{eq:KM_Cartan}
&& \{ \Hc_0^i \,, \hat k \,, L_0 \} \,, \quad i = 1\,, ... \,, r  \,,
\eeqn
while all other generators of $\{ E_n^{ \boldsymbol{\alpha }}  \}$ and $\{ \Hc_n^i \}$ (with $n\neq 0$) form the ladder operators.

\para
The affine roots include those associated with the ladder operators of ${E^{ \boldsymbol{\alpha } } }_n$
\beqn\label{eq:affine_roots}
\hat{ \boldsymbol{ \alpha } } &\equiv& ( \boldsymbol{ \alpha } \,; 0 \,; n) = \boldsymbol{ \alpha }  + n \boldsymbol{ \delta } \,, \quad \boldsymbol{ \alpha } \in \Delta\,, \quad  \boldsymbol{ \delta } =( \boldsymbol{ 0 } \,; 0\,; 1 ) \,, \quad n \in \mathbb{Z}  \,,
\eeqn
as well as an extra simple root of
\beqn\label{eq:affine_root0}
\boldsymbol{ \alpha }_0  &\equiv& ( - \boldsymbol{ \theta } \,; 0 \,; 1)= -  \boldsymbol{ \theta } +  \boldsymbol{ \delta } \,.
\eeqn
Here, the $( \boldsymbol{ \alpha } \,; 0 \,; 0)$ represents the finite part of the affine roots $\hat{ \boldsymbol{ \alpha } }$, $\boldsymbol{ \theta }$ is the highest root of $\gG$ in Eq.~\eqref{eq:highest_root}, and the $\boldsymbol{ \delta }$ is known as the imaginary root due to its zero length of $( \boldsymbol{ \delta } \,, \boldsymbol{ \delta }  )=0$.

\para
The affine weights are generally denoted as
\beqn
\hat{ \boldsymbol{\lambda } }&=& (  \boldsymbol{\lambda }  \,; k_{ \boldsymbol{\lambda } } \,; n_{ \boldsymbol{\lambda } } )\,,
\eeqn
with the scalar products defined by
\beqn
( \hat{ \boldsymbol{\lambda } } \,, \hat{ \boldsymbol{\sigma } } )&\equiv& ( \boldsymbol{\lambda } \,, \boldsymbol{\sigma } ) + k_{ \boldsymbol{\lambda }} n_{ \boldsymbol{\sigma }} + k_{ \boldsymbol{\sigma }} n_{ \boldsymbol{\lambda }}  \,.
\eeqn
The fundamental weights of the $\hat \gG$ is given by
\beqn
&& \hat{ \boldsymbol{\omega } }_i \equiv a_i^\lor \hat{ \boldsymbol{\omega } }_0 + ( \boldsymbol{\omega }_i \,; 0 \,; 0 ) \,,\quad  \hat{ \boldsymbol{\omega } }_0 = ( \boldsymbol{0} \,; 1 \,; 0 ) \,,
%\begin{cases}
%(0\,; 1\,; 0 )\,, \mbox{ if $i=0$} \,,\\
%\,, \mbox{ if $i\neq 0$}
%\end{cases}
\eeqn
with the $\boldsymbol{0}$ representing the $N-1$ zeros in the {\it basic fundamental weight} of $\hat{ \boldsymbol{\omega } }_0$.
The corresponding scalar products are given by
\beqn
&& ( \hat{ \boldsymbol{\omega } }_i \,, \hat{ \boldsymbol{\omega } }_0 )  = ( \hat{ \boldsymbol{\omega } }_0 \,, \hat{ \boldsymbol{\omega } }_0 ) = 0\,, \quad ( \hat{ \boldsymbol{\omega } }_i \,, \hat{ \boldsymbol{\omega } }_j ) = (  \boldsymbol{\omega }_i \,,  \boldsymbol{\omega }_j ) = F_{ij} \,,\quad (i\,, j \neq 0) \,.
\eeqn
For the $\gG=\sG\uG(N)$, the quadratic form matrix of $F_{ij}$ was previously given in Eq.~\eqref{eq:quadratic_form}.
Any affine weight can be expanded in terms of the fundamental weights $\hat{ \boldsymbol{\omega } }_i$ and the imaginary root $\boldsymbol{\delta }$ as
\beqn
\hat{ \boldsymbol{\lambda } } &=&  \sum_{ i=0 }^r \lambda_i  \hat{ \boldsymbol{\omega } }_i + \ell \boldsymbol{\delta } \,, \quad \ell\in \mathbb{R} \,,
\eeqn
with $\ell$ being the {\it grade}.
For the affine Lie algebra of $\widehat{ \sG \uG}(N)_{k}$, the affine level is given by
\beqn\label{eq:affine_level}
k&\equiv&  \hat{ \boldsymbol{\lambda } } ( \hat k ) = (  \hat{ \boldsymbol{\lambda } } \,, \boldsymbol{\delta } )= \sum_{i=0}^r \lambda_i \,.
\eeqn
with $a_i^\lor=1$ for the $\sG\uG(N)$.
The affine weights are also expressed in terms of the Dynkin labels as follows
\beqn
&& \hat{ \boldsymbol{\lambda } } = [  \lambda_0 \,; \lambda_1 \,, ... \,, \lambda_r  ] \,.
\eeqn
For all $\widehat{ \sG\uG}(8)_{k_U=1}$ rank-$k$ anti-symmetric irreps, their affine weights can still be expressed as the following highest-weight states
\beqn
\hat{ \boldsymbol{\lambda } } ( A_{k\, {\tt HW}} )&=& [ 0\,;  \underbrace{ 0\,, ... \,,0 }_{k-1} \,,1\,, \underbrace{ 0 \,, ... \,, 0}_{ 7-k } ] \,.
\eeqn
Previously, it was argued~\cite{Dienes:1996du} that the adjoint representation cannot be admitted in the $k=1$ affine Lie algebra since one would naively express it in terms of the highest-weight state as
\beqn
\hat{ \boldsymbol{\lambda } } ( {\rm adj}_{\tt HW} )&=& [ 0\,; 1 \,, \underbrace{ 0\,, ... \,,0 }_{ 5} \,,1 ] \,.
\eeqn
Instead, it can be obtained by subtracting a new simple root in Eq.~\eqref{eq:affine_root0} from the $\widehat{ \sG\uG}(8)_{k_U=1}$ singlet (and highest-weight) state as follows
\beqn\label{eq:Adj_descendant}
\hat{ \boldsymbol{\lambda } } ( {\rm adj}_{\tt des} )&=& ( \boldsymbol{ 0 } \,; 1 \,; 0 ) - \boldsymbol{\alpha }_0 = ( \boldsymbol{ \theta } \,; 1 \,; -1 ) \,.
%\hat{ \boldsymbol{\lambda } } ( {\rm adj} )&=& [ -1\,; 1 \,, \underbrace{ 0\,, ... \,,0 }_{ 5} \,,1 ] = [ 1\,;  \underbrace{ 0\,, ... \,,0 }_{  7 }  ] - \,,
\eeqn
This can be achieved with the fact that both the adjoint representation and the singlet representation live in the same congruency class since
\beqn
&&  \Big( \sum_{i=1}^7 i \cdot \lambda_i \Big) [   \underbrace{ 0\,, ... \,,0 }_{7}  ] =  \Big( \sum_{i=1}^7 i \cdot \lambda_i \Big) [   1\,, \underbrace{ 0\,, ... \,,0 }_{5 } \,,1 ]  = 0 ~{\rm mod}~ 8\,.
\eeqn
In other words, the adjoint representation should be interpreted as the descendant state from the $\widehat{ \sG\uG}(8)_{k_U=1}$ singlet state of $( \boldsymbol{ 0 } \,; 1 \,; 0 )$.
The last entry of $-1$ in Eq.~\eqref{eq:Adj_descendant} represents a grade$=1$ state, where the minus sign originates from the definition of the grading operator in Eq.~\eqref{eq:grading_Op}.
Therefore, we find the conformal dimension for the adjoint representation as
\beqn
&& h( {\rm adj}_{\tt des} ) = h (  \boldsymbol{ 0 } \,; 1 \,; 0  ) + {\rm grade} = 1\,,
\eeqn
which apparently satisfies the unitarity constraint in Eq.~\eqref{eq:conformal_unitarity}.
One seemingly possible SUSY extension in Eq.~\eqref{eq:SUSY_SU8_chiralsII} involves two chiral superfields of $\rep{36_H}^\rho$.
This rank-$2$ symmetric irrep cannot be understood as the highest-weight state, since
\beqn
\hat{ \boldsymbol{\lambda } } ( \rep{36}_{\tt HW}  )&=& [ 0\,; 2 \,, \underbrace{ 0\,, ... \,,0 }_{ 6}  ] \Rightarrow k( \rep{36}_{\tt HW}  ) = 2 \,,
\eeqn
according to Eq.~\eqref{eq:affine_level}.
Instead, it can only be a $k_U=1$ and grade$=1$ descendant state obtained from the highest-weight state of $\rep{28}_{\rm HW}$ as follows
\beqn
&& \hat{ \boldsymbol{\lambda } } ( \rep{36}_{\tt des}  )= [ -1\,; 2 \,, \underbrace{ 0\,, ... \,,0 }_{ 6}  ]  = \hat{ \boldsymbol{\lambda } }  ( \rep{28}_{\tt HW}) -  ( \sum_{i=2}^7 \boldsymbol{\alpha }_i  ) - \boldsymbol{\alpha }_0 \,, \non
{\rm with}&&~ \hat{ \boldsymbol{\lambda } }  ( \rep{28}_{\tt HW}) = [ 0\,; 0\,, 1 \,, \underbrace{ 0\,, ... \,,0 }_{ 5}  ] \,,
\eeqn
since the $\rep{36}$ and the $\rep{28}$ also live in the same congruency class as
\beqn
&&  \Big( \sum_{i=1}^7 i \cdot \lambda_i \Big) [   2\,, \underbrace{ 0\,, ... \,,0 }_{6}  ] =  \Big( \sum_{i=1}^7 i \cdot \lambda_i \Big) [  0\,, 1\,, \underbrace{ 0\,, ... \,,0 }_{5 }  ]  = 2 ~{\rm mod}~ 8\,.
\eeqn
Therefore, the conformal dimension for the $\rep{36}$ representation as the descendant state violates the unitarity constraint in Eq.~\eqref{eq:conformal_unitarity} as follows
\beqn
&& h( \rep{36}_{\tt des} ) = h (  \rep{28}_{\tt HW} ) + {\rm grade} = \frac{3}{4}+1 = \frac{7 }{4} >1 \,.
\eeqn
It means the naive SUSY extension in Eq.~\eqref{eq:SUSY_SU8_chiralsII} cannot be realized if one hypothesizes the corresponding affine Lie algebra of $\widehat{ \sG\uG}(8)_{ k_U =1}$.

%###################################################################
\section{The ${\uG}(1)$ Cartan discontinuities}
\label{section:discont}
%###################################################################

\para
In the class of GUTs beyond the ${\sG\uG}(5)$, one usually has the intermediate symmetry breaking stages of the form
\beqn
&& {\sG\uG}(N) \oplus {\uG}(1)_X  \hookrightarrow  {\sG\uG}(N-1) \oplus  {\uG}(1)_{X^\prime } \,,
\eeqn
where the corresponding gauge couplings are denoted as
\beqn
&& ( g_N \,, g_X ) \,,\quad  ( g_{N-1} \,, g_{X^\prime} )\,,
\eeqn
respectively.
Such a symmetry breaking pattern is usually achieved by the Higgs fields of $\rep{\Phi} \equiv ( \rep{N} \,, +\frac{1}{N} )_{ \mathbf{H} } \in {\sG\uG}(N) \oplus {\uG}(1)_X$, whose decomposition and the VEV can be denoted as follows
\beqn\label{eq:SUN_decomp}
&& ( \rep{N} \,, +\frac{1}{N} )_{ \mathbf{H} }  \hookrightarrow  ( \rep{N-1} \,, +\frac{1}{N-1} )_{ \mathbf{H} } \oplus ( \rep{1} \,, 0 )_{ \mathbf{H} } \,, \non
&& \langle  \rep{\Phi}  \rangle = \frac{1}{ \sqrt{2} } ( \vec{0}_{N-1} \,, v_N )^T \,.
\eeqn
To find the relation between two ${\uG}(1)$ gauge couplings of $g_X$ and $g_{X^\prime}$, we extrapolate the following gauge boson mass squared terms from the covariant derivative
\beqn\label{eq:SUN_cov}
| D_\mu \rep{\Phi } |^2 &\supset& \frac{ v_N^2 }{2 } (  - g_N \sqrt{ \frac{ N-1 }{ 2N } }W_\mu^{ N^2 -1 } + \frac{ g_X }{ N }  X_\mu )^2 \,,
\eeqn
where only the component associated with the last ${\sG\uG}(N)$ Cartan generator of $T_{ {\sG\uG}(N) }^{ N^2 -1}$ is necessary.
Thus, we find the following mixing between gauge bosons of $( W_\mu^{ N^2 -1 } \,, X_\mu)$
\beqn\label{eq:SUN_GBmix}
\left( \ba{c}  
W_\mu^{ N^2 -1 }   \\
X_\mu   \\  \ea  \right)  &=& \left( \ba{cc}  
\cos \theta_N &  \sin \theta_N    \\
- \sin \theta_N   & \cos \theta_N   \\  \ea  \right)   \cdot  \left( \ba{c}  
Z_\mu^{ N -1 }   \\
X_\mu^\prime   \\  \ea  \right)   \,, \quad   \tan\theta_N = \frac{ g_X }{ g_N } \sqrt{  \frac{2}{  N (N - 1 ) } } \,,
\eeqn
where the $Z_\mu^{ N -1 }$ is massive and the $X_\mu^\prime$ is massless.
By expressing the term in Eq.~\eqref{eq:SUN_cov} with the mixing relation in Eq.~\eqref{eq:SUN_GBmix}, we have to require the $X_\mu^\prime $ gauge coupling to match with the decomposition in Eq.~\eqref{eq:SUN_decomp}
\beqn
&&  g_N \sqrt{ \frac{ 1 }{ 2N ( N-1) } }W_\mu^{ N^2 -1 } + \frac{g_X}{ N }  X_\mu  \non
&\supset&  ( \frac{ g_N}{ \sqrt{ 2N (N-1) } } \sin\theta_N  + \frac{ g_X}{N} \cos\theta_N ) X_\mu^\prime = \frac{ g_{X^\prime}}{ N-1} X_\mu^\prime \,,
\eeqn
which leads to the general ${\uG}(1)$ Cartan discontinuities of
\beqn\label{eq:Cartan_discont}
&& \alpha_{ X^\prime }^{-1} =  \frac{2}{ N(N-1)} \alpha_{N}^{-1} +  \alpha_{X}^{-1} \,.
\eeqn
Obviously, this relation originates from the last ${\sG\uG}(N)$ Cartan generator of $T_{ {\sG\uG}(N) }^{ N^2 -1}$.
Specifically, we have the relations of
\beqs
\beqn
{\sG\uG}(4) \oplus {\uG}(1)_X  \hookrightarrow  {\sG\uG}(3) \oplus {\uG}(1)_{X^\prime }~&:&~  \alpha_{ X^\prime }^{-1} =  \frac{ 1 }{ 6 } \alpha_{4}^{-1} +  \alpha_{X}^{-1} \,, \\[1mm]
{\sG\uG}(3) \oplus {\uG}(1)_X  \hookrightarrow  {\sG\uG}(2) \oplus {\uG}(1)_{X^\prime } ~&:&~ \alpha_{ X^\prime }^{-1} =  \frac{ 1 }{ 3 } \alpha_{ 3}^{-1} +  \alpha_{X}^{-1} \,,
\eeqn
\eeqs
as were previously displayed in Figs.~\ref{fig:RGE_mini} and \ref{fig:RGE_TypeI}.

%\bibliographystyle{utphys.bst}
%\bibliography{references}

\begin{thebibliography}{10}

\bibitem{Georgi:1974sy}
H.~Georgi and S.~L. Glashow, ``{Unity of All Elementary Particle Forces},''
  \href{http://dx.doi.org/10.1103/PhysRevLett.32.438}{{\em Phys. Rev. Lett.}
  {\bfseries 32} (1974) 438--441}.

\bibitem{Fritzsch:1974nn}
H.~Fritzsch and P.~Minkowski, ``{Unified Interactions of Leptons and
  Hadrons},'' \href{http://dx.doi.org/10.1016/0003-4916(75)90211-0}{{\em Annals
  Phys.} {\bfseries 93} (1975) 193--266}.

\bibitem{Cabibbo:1963yz}
N.~Cabibbo, ``{Unitary Symmetry and Leptonic Decays},''
  \href{http://dx.doi.org/10.1103/PhysRevLett.10.531}{{\em Phys. Rev. Lett.}
  {\bfseries 10} (1963) 531--533}.

\bibitem{Kobayashi:1973fv}
M.~Kobayashi and T.~Maskawa, ``{CP Violation in the Renormalizable Theory of
  Weak Interaction},'' \href{http://dx.doi.org/10.1143/PTP.49.652}{{\em Prog.
  Theor. Phys.} {\bfseries 49} (1973) 652--657}.

\bibitem{Georgi:1979md}
H.~Georgi, ``{Towards a Grand Unified Theory of Flavor},''
  \href{http://dx.doi.org/10.1016/0550-3213(79)90497-8}{{\em Nucl. Phys. B}
  {\bfseries 156} (1979) 126--134}.

\bibitem{CMS:2022dwd}
{\bfseries CMS} Collaboration, A.~Tumasyan {\em et~al.}, ``{A portrait of the
  Higgs boson by the CMS experiment ten years after the discovery.},''
  \href{http://dx.doi.org/10.1038/s41586-022-04892-x}{{\em Nature} {\bfseries
  607} no.~7917, (2022) 60--68},
  \href{http://arxiv.org/abs/2207.00043}{{\ttfamily arXiv:2207.00043
  [hep-ex]}}.

\bibitem{ATLAS:2022vkf}
{\bfseries ATLAS} Collaboration, G.~Aad {\em et~al.}, ``{A detailed map of
  Higgs boson interactions by the ATLAS experiment ten years after the
  discovery},'' \href{http://dx.doi.org/10.1038/s41586-022-04893-w}{{\em
  Nature} {\bfseries 607} no.~7917, (2022) 52--59},
  \href{http://arxiv.org/abs/2207.00092}{{\ttfamily arXiv:2207.00092
  [hep-ex]}}. [Erratum: Nature 612, E24 (2022)].

\bibitem{Chen:2023qxi}
N.~Chen, Y.-n. Mao, and Z.~Teng, ``{The global B \ensuremath{-} L symmetry in
  the flavor-unified SU(N) theories},''
  \href{http://dx.doi.org/10.1007/JHEP04(2024)046}{{\em JHEP} {\bfseries 04}
  (2024) 046}, \href{http://arxiv.org/abs/2307.07921}{{\ttfamily
  arXiv:2307.07921 [hep-ph]}}.

\bibitem{Barr:2008pn}
S.~M. Barr, ``{Doubly Lopsided Mass Matrices from Unitary Unification},''
  \href{http://dx.doi.org/10.1103/PhysRevD.78.075001}{{\em Phys. Rev. D}
  {\bfseries 78} (2008) 075001},
  \href{http://arxiv.org/abs/0804.1356}{{\ttfamily arXiv:0804.1356 [hep-ph]}}.

\bibitem{Chen:2024cht}
N.~Chen, Y.-n. Mao, and Z.~Teng, ``{The Standard Model quark/lepton masses and
  the Cabibbo-Kobayashi-Maskawa mixing in an ${\rm SU}(8)$ theory},''
  \href{http://arxiv.org/abs/2402.10471}{{\ttfamily arXiv:2402.10471
  [hep-ph]}}.

\bibitem{Peccei:1977hh}
R.~D. Peccei and H.~R. Quinn, ``{CP Conservation in the Presence of
  Instantons},'' \href{http://dx.doi.org/10.1103/PhysRevLett.38.1440}{{\em
  Phys. Rev. Lett.} {\bfseries 38} (1977) 1440--1443}.

\bibitem{Li:1973mq}
L.-F. Li, ``{Group Theory of the Spontaneously Broken Gauge Symmetries},''
  \href{http://dx.doi.org/10.1103/PhysRevD.9.1723}{{\em Phys. Rev. D}
  {\bfseries 9} (1974) 1723--1739}.

\bibitem{Chen:2024deo}
N.~Chen, Z.~Hou, Y.-n. Mao, and Z.~Teng, ``{The gauge coupling evolutions of an
  SU(8) theory with the maximally symmetry breaking pattern},''
  \href{http://dx.doi.org/10.1007/JHEP10(2024)149}{{\em JHEP} {\bfseries 10}
  (2024) 149}, \href{http://arxiv.org/abs/2406.09970}{{\ttfamily
  arXiv:2406.09970 [hep-ph]}}.

\bibitem{Chen:2024yhb}
N.~Chen, Z.~Chen, Z.~Hou, Z.~Teng, and B.~Wang, ``{Further study of the
  maximally symmetry breaking patterns in an ${\rm SU}(8)$ theory},''
  \href{http://arxiv.org/abs/2409.03172}{{\ttfamily arXiv:2409.03172
  [hep-ph]}}.

\bibitem{Georgi:1974yf}
H.~Georgi, H.~R. Quinn, and S.~Weinberg, ``{Hierarchy of Interactions in
  Unified Gauge Theories},''
  \href{http://dx.doi.org/10.1103/PhysRevLett.33.451}{{\em Phys. Rev. Lett.}
  {\bfseries 33} (1974) 451--454}.

\bibitem{Hall:1980kf}
L.~J. Hall, ``{Grand Unification of Effective Gauge Theories},''
  \href{http://dx.doi.org/10.1016/0550-3213(81)90498-3}{{\em Nucl. Phys. B}
  {\bfseries 178} (1981) 75--124}.

\bibitem{Dimopoulos:1981zb}
S.~Dimopoulos and H.~Georgi, ``{Softly Broken Supersymmetry and SU(5)},''
  \href{http://dx.doi.org/10.1016/0550-3213(81)90522-8}{{\em Nucl. Phys. B}
  {\bfseries 193} (1981) 150--162}.

\bibitem{Weinberg:1980wa}
S.~Weinberg, ``{Effective Gauge Theories},''
  \href{http://dx.doi.org/10.1016/0370-2693(80)90660-7}{{\em Phys. Lett. B}
  {\bfseries 91} (1980) 51--55}.

\bibitem{Hill:1983xh}
C.~T. Hill, ``{Are There Significant Gravitational Corrections to the
  Unification Scale?},''
  \href{http://dx.doi.org/10.1016/0370-2693(84)90451-9}{{\em Phys. Lett. B}
  {\bfseries 135} (1984) 47--51}.

\bibitem{Shafi:1983gz}
Q.~Shafi and C.~Wetterich, ``{Modification of {GUT} Predictions in the Presence
  of Spontaneous Compactification},''
  \href{http://dx.doi.org/10.1103/PhysRevLett.52.875}{{\em Phys. Rev. Lett.}
  {\bfseries 52} (1984) 875}.

\bibitem{Glashow:1979nm}
S.~L. Glashow, ``{The Future of Elementary Particle Physics},''
  \href{http://dx.doi.org/10.1007/978-1-4684-7197-7_15}{{\em NATO Sci. Ser. B}
  {\bfseries 61} (1980) 687}.

\bibitem{Barbieri:1979ag}
R.~Barbieri, D.~V. Nanopoulos, G.~Morchio, and F.~Strocchi, ``{Neutrino Masses
  in Grand Unified Theories},''
  \href{http://dx.doi.org/10.1016/0370-2693(80)90058-1}{{\em Phys. Lett. B}
  {\bfseries 90} (1980) 91--97}.

\bibitem{Barbieri:1980vc}
R.~Barbieri and D.~V. Nanopoulos, ``{An Exceptional Model for Grand
  Unification},'' \href{http://dx.doi.org/10.1016/0370-2693(80)90998-3}{{\em
  Phys. Lett. B} {\bfseries 91} (1980) 369--375}.

\bibitem{Barbieri:1980tz}
R.~Barbieri and D.~V. Nanopoulos, ``{Hierarchical Fermion Masses From Grand
  Unification},'' \href{http://dx.doi.org/10.1016/0370-2693(80)90395-0}{{\em
  Phys. Lett. B} {\bfseries 95} (1980) 43--46}.

\bibitem{delAguila:1980qag}
F.~del Aguila and L.~E. Ibanez, ``{Higgs Bosons in SO(10) and Partial
  Unification},'' \href{http://dx.doi.org/10.1016/0550-3213(81)90266-2}{{\em
  Nucl. Phys. B} {\bfseries 177} (1981) 60--86}.

\bibitem{Ginsparg:1987ee}
P.~H. Ginsparg, ``{Gauge and Gravitational Couplings in Four-Dimensional String
  Theories},'' \href{http://dx.doi.org/10.1016/0370-2693(87)90357-1}{{\em Phys.
  Lett. B} {\bfseries 197} (1987) 139--143}.

\bibitem{Font:1990uw}
A.~Font, L.~E. Ibanez, and F.~Quevedo, ``{Higher Level {Kac-Moody} String
  Models and Their Phenomenological Implications},''
  \href{http://dx.doi.org/10.1016/0550-3213(90)90393-R}{{\em Nucl. Phys. B}
  {\bfseries 345} (1990) 389--430}.

\bibitem{Dienes:1996du}
K.~R. Dienes, ``{String theory and the path to unification: A Review of recent
  developments},'' \href{http://dx.doi.org/10.1016/S0370-1573(97)00009-4}{{\em
  Phys. Rept.} {\bfseries 287} (1997) 447--525},
  \href{http://arxiv.org/abs/hep-th/9602045}{{\ttfamily arXiv:hep-th/9602045}}.

\bibitem{Goddard:1986bp}
P.~Goddard and D.~I. Olive, ``{Kac-Moody and Virasoro Algebras in Relation to
  Quantum Physics},'' \href{http://dx.doi.org/10.1142/S0217751X86000149}{{\em
  Int. J. Mod. Phys. A} {\bfseries 1} (1986) 303}.

\bibitem{Machacek:1983tz}
M.~E. Machacek and M.~T. Vaughn, ``{Two Loop Renormalization Group Equations in
  a General Quantum Field Theory. 1. Wave Function Renormalization},''
  \href{http://dx.doi.org/10.1016/0550-3213(83)90610-7}{{\em Nucl. Phys. B}
  {\bfseries 222} (1983) 83--103}.

\bibitem{Antoniadis:1982vr}
I.~Antoniadis, C.~Kounnas, and K.~Tamvakis, ``{Simple Treatment of Threshold
  Effects},'' \href{http://dx.doi.org/10.1016/0370-2693(82)90693-1}{{\em Phys.
  Lett. B} {\bfseries 119} (1982) 377--380}.

\bibitem{Langacker:1992rq}
P.~Langacker and N.~Polonsky, ``{Uncertainties in coupling constant
  unification},'' \href{http://dx.doi.org/10.1103/PhysRevD.47.4028}{{\em Phys.
  Rev. D} {\bfseries 47} (1993) 4028--4045},
  \href{http://arxiv.org/abs/hep-ph/9210235}{{\ttfamily arXiv:hep-ph/9210235}}.

\bibitem{Martin:1993zk}
S.~P. Martin and M.~T. Vaughn, ``{Two loop renormalization group equations for
  soft supersymmetry breaking couplings},''
  \href{http://dx.doi.org/10.1103/PhysRevD.50.2282}{{\em Phys. Rev. D}
  {\bfseries 50} (1994) 2282},
  \href{http://arxiv.org/abs/hep-ph/9311340}{{\ttfamily arXiv:hep-ph/9311340}}.
  [Erratum: Phys.Rev.D 78, 039903 (2008)].

\bibitem{Witten:1981nf}
E.~Witten, ``{Dynamical Breaking of Supersymmetry},''
  \href{http://dx.doi.org/10.1016/0550-3213(81)90006-7}{{\em Nucl. Phys. B}
  {\bfseries 188} (1981) 513}.

\bibitem{DiFrancesco:1997nk}
P.~Di~Francesco, P.~Mathieu, and D.~Senechal,
  \href{http://dx.doi.org/10.1007/978-1-4612-2256-9}{{\em {Conformal Field
  Theory}}}.
\newblock Graduate Texts in Contemporary Physics. Springer-Verlag, New York,
  1997.

\end{thebibliography}

\providecommand{\href}[2]{#2}\begingroup\raggedright\endgroup

\end{document}